\documentclass{ieeeaccess}
\usepackage{cite}
\usepackage{amsmath,amssymb,amsfonts}
\usepackage{algorithmic}
\usepackage{graphicx}
\usepackage{textcomp}

\usepackage{bm}
\makeatletter
\AtBeginDocument{\DeclareMathVersion{bold}
\SetSymbolFont{operators}{bold}{T1}{times}{b}{n}
\SetSymbolFont{NewLetters}{bold}{T1}{times}{b}{it}
\SetMathAlphabet{\mathrm}{bold}{T1}{times}{b}{n}
\SetMathAlphabet{\mathit}{bold}{T1}{times}{b}{it}
\SetMathAlphabet{\mathbf}{bold}{T1}{times}{b}{n}
\SetMathAlphabet{\mathtt}{bold}{OT1}{pcr}{b}{n}
\SetSymbolFont{symbols}{bold}{OMS}{cmsy}{b}{n}
\renewcommand\boldmath{\@nomath\boldmath\mathversion{bold}}}
\makeatother

\def\BibTeX{{\rm B\kern-.05em{\sc i\kern-.025em b}\kern-.08em
    T\kern-.1667em\lower.7ex\hbox{E}\kern-.125emX}}

\usepackage{siunitx}
\DeclareSIUnit{\flop}{\text{Flop}}
\DeclareSIUnit{\flops}{\text{Flop}/\text{s}}
\DeclareSIUnit{\byte}{\text{B}}
\DeclareSIUnit{\bytes}{\text{B}/\text{s}}

\usepackage{tabularray}
\usepackage{rotating}
\usepackage{url}
\RequirePackage[l2tabu, orthodox]{nag}
\usepackage[all, warning]{onlyamsmath}

\usepackage{derivative}
\usepackage{physics2}
\usephysicsmodule{ab}

\usepackage{xcolor}
\definecolor{UDred}{HTML}{ff4b00}
\definecolor{UDblue}{HTML}{005aff}
\definecolor{UDorange}{HTML}{f6aa00}
\definecolor{UDgreen}{HTML}{03af7a}
\definecolor{UDcyan}{HTML}{4dc4ff}
\definecolor{UDbrown}{HTML}{804000}
\definecolor{UDpurple}{HTML}{990099}
\definecolor{UDyellow}{HTML}{fff100}
\definecolor{UDpink}{HTML}{ff8082}

\usepackage{listings}
\usepackage{seqsplit}
\usepackage{fancyvrb}
\VerbatimFootnotes

\makeatletter
\newif\ifreview@ba@break
\def\review@ba@end{\review@ba@end@}
\DeclareRobustCommand{\reviewbreakall}[1]{%
  \begingroup
    \review@ba@breakfalse
    \review@break@all@a#1\review@ba@end
  \endgroup
}
\def\review@break@all@a{%
  \futurelet\review@ba@tok\review@break@all@b
}
\def\review@break@all@b{%
  \ifx\review@ba@tok\review@ba@end
    \let\next\@gobble
  \else\ifx\review@ba@tok\@sptoken
    \let\next\review@break@all@c
  \else\ifx\review@ba@tok~%
    \let\next\review@break@all@d
  \else\ifx\review@ba@tok\bgroup
    \let\next\review@break@all@e
  \else
    \let\next\review@break@all@f
  \fi\fi\fi\fi
  \next
}
\expandafter\def\expandafter\review@break@all@c\space{%
  \space
  \review@ba@breakfalse
  \review@break@all@a
}
\def\review@break@all@d#1{%
  \review@break@all@f{\mbox{\space}}%
}
\def\review@break@all@e#1{%
  \review@break@all@f{{#1}}%
}
\def\review@break@all@f#1{%
  \ifreview@ba@break
    \hskip0pt plus 0.02em\relax
  \fi
  #1%
  \review@ba@breaktrue
  \review@break@all@a
}

\DeclareRobustCommand{\reviewtt}[1]{{\ttfamily\reviewbreakall{#1}}}

\makeatother

\newcommand{\Hopper}{NVIDIA H100 SXM 80GB}
\newcommand{\GraceHopper}{NVIDIA GH200 480GB}
\newcommand{\Aldebaran}{AMD Instinct MI210}
\newcommand{\PVC}{Intel Data Center GPU Max 1100}
\newcommand{\HopperShort}{NVIDIA H100}
\newcommand{\GraceHopperShort}{NVIDIA GH200}
\newcommand{\AldebaranShort}{AMD MI210}
\newcommand{\PVCshort}{Intel PVC1100}

\newcommand{\acppFull}{AdaptiveCpp}
\newcommand{\icpxFull}{Intel oneAPI}

\newcommand{\LibName}{Solomon}

\begin{document}
\history{Date of publication xxxx 00, 0000, date of current version xxxx 00, 0000.}
\doi{10.1109/ACCESS.2024.0429000}

\title{Unified schemes for directive-based GPU offloading}
\author{\uppercase{Yohei Miki}\authorrefmark{1}, \IEEEmembership{Member, IEEE},
and \uppercase{Toshihiro Hanawa}\authorrefmark{1},\IEEEmembership{Member, IEEE}}

\address[1]{Information Technology Center, The University of Tokyo, 6-2-3 Kashiwanoha, Kashiwa, Chiba 277-0882, Japan}

\markboth
{Miki \& Hanawa: \LibName{} (Simple Off-LOading Macros Orchestrating multiple Notations)}
{Miki \& Hanawa: \LibName{} (Simple Off-LOading Macros Orchestrating multiple Notations)}

\corresp{Corresponding author: Yohei Miki (e-mail: ymiki@cc.u-tokyo.ac.jp).}

\begin{abstract}
    GPU is the dominant accelerator device due to its high performance and energy efficiency.
    Directive-based GPU offloading using OpenACC or OpenMP target is a convenient way to port existing codes originally developed for multicore CPUs.
    Although OpenACC and OpenMP target provide similar features, both methods have pros and cons.
    OpenACC has better functions and an abundance of documents, but it is virtually for NVIDIA GPUs.
    OpenMP target supports NVIDIA/AMD/Intel GPUs but has fewer functions than OpenACC.
    Here, we have developed a header-only library, \LibName{} (Simple Off-LOading Macros Orchestrating multiple Notations), to unify the interface for GPU offloading with the support of both OpenACC and OpenMP target.
    \LibName{} provides three types of notations to reduce users' implementation and learning costs: intuitive notation for beginners and OpenACC/OpenMP-like notations for experienced developers.
    This manuscript denotes \LibName{}'s implementation and usage and demonstrates the GPU-offloading in $N$-body simulation and the three-dimensional diffusion equation.
    The library and sample codes are provided as open-source software and publicly and freely available at \url{https://github.com/ymiki-repo/solomon}.
\end{abstract}

\begin{keywords}
    directive, GPU, OpenACC, OpenMP target, preprocessor macro, vendor lock-in
\end{keywords}

\titlepgskip=-21pt

\maketitle

\section{Introduction}
\label{sec:introduction}
\renewcommand{\thefootnote}{\fnsymbol{footnote}}
\footnote[0]{\copyright{} 2024 IEEE. Personal use of this material is permitted. Permission from IEEE must be obtained for all other uses, in any current or future media, including reprinting/republishing this material for advertising or promotional purposes, creating new collective works, for resale or redistribution to servers or lists, or reuse of any copyrighted component of this work in other works.}
\renewcommand{\thefootnote}{\arabic{footnote}}
Compute demands have continuously increased, and simultaneously, pressure to reduce electricity for computation has also been strengthened.
These contradictory requirements offer a high-performance and energy-efficient device.
GPUs meet the demand, so over $1/3$ of supercomputers listed in the latest TOP500 list equipped GPUs as accelerator devices \cite{TOP500November2024}.
Most GPU-accelerated supercomputers have relied on NVIDIA GPUs.
However, the GPU vendors' diversification is ongoing; for example, AMD and Intel GPUs boost the exascale supercomputers (Frontier and Aurora, respectively).

The diversification of the GPU vendors causes the diversification of the programming environments in GPU computing \cite{Herten2023}.
CUDA/HIP/SYCL is the first choice in the performance optimizations specialized for NVIDIA/AMD/Intel GPUs \cite{MikiHanawa2024}.
Although such dedicated optimizations provide the highest performance in each environment, enormous porting efforts from the original CPU code sometimes hamper the developers' adoption.
The more convenient GPU-programming method is directive-based programming using OpenACC \cite{OpenACC2.7} or OpenMP target directives \cite{OpenMP5.2}, which is frequently adopted, especially in the GPU porting of large codes.

The rich functions and abundance of documents due to the first-mover advantage push OpenACC as the first choice for directive-based GPU porting.
Better coordination of asynchronous execution of multiple operations is the representative benefit of OpenACC over OpenMP target.
In OpenMP target, \texttt{nowait} and \texttt{taskwait} enable asynchronous features.
The corresponding indications in OpenACC are \texttt{async [(expression)]} and \texttt{wait [(expression)]}.
The additional input of ``\texttt{(expression)}'' is helpful to tag multiple queues; therefore, OpenACC can support more sophisticated control of multiple queues than OpenMP target.
However, the acquisition of PGI by NVIDIA made OpenACC virtually for NVIDIA GPUs.
The HPE Cray compiler provided by HPE support OpenACC-offloading for AMD GPUs, relieving vendor lock-in.
Note that the compiler is bundled in the HPE-provided supercomputers, so vendor lock-in still exists.
On the other hand, OpenMP target directives support all major GPU vendors but still lack a part of the functions provided in OpenACC.
Furthermore, the additional learning costs burden experienced developers in OpenACC to learn and adopt OpenMP target directives.

In this study, we have developed a header-only library, \LibName{} (Simple Off-LOading Macros Orchestrating multiple Notations), to provide a directive-like programming interface supporting both OpenACC and OpenMP target as backends.
\LibName{} enables a single code run using OpenACC on NVIDIA GPU or OpenMP target on NVIDIA/AMD/Intel GPUs.
Moreover, \LibName{} provides OpenACC and OpenMP-like notations to reduce the learning cost of experienced developers.
Since \LibName{} is just an aggregate of preprocessor macros corresponding to directives and clauses provided in OpenACC or OpenMP, users can use compilers provided by GPU-manufacturing vendors as is.
It means that users will benefit directly from the improvement of the vendors' compilers.

\LibName{} resembles HIP developed by AMD in the feature, realizing a code developed in a unified interface can run on GPUs by multiple vendors.
HIP is a CUDA-like programming environment and uses CUDA as the backend compiler on NVIDIA GPU; therefore, there is no penalty in performance and function \cite{AMD_HIP}.
\LibName{} is like a directive-version HIP but is the header-only library without compiler-like features, meaning installation is effortless in any environment.
SYCL \cite{SYCL} and OCCA \cite{OCCA} also support multiple device APIs by a unified interface, but the learning costs are not minimized.
Another favored method for realizing portability among multiple vendor GPUs is performance-portable frameworks like Kokkos \cite{CarterEdwards2014,Trott2022}.
Kokkos can use CUDA, HIP, and SYCL as backends those are lower-level programming models than directives; therefore, Kokkos would provide better performance than directive-based implementations.
On the other hand, Kokkos offers substantial porting costs from the original CPU implementation to exploit Kokkos API (lambda expression in many cases), and the software's readability decreases compared with directives.
Directive-like interfaces adopted in \LibName{} ease the porting from existing codes, keeping the implementation readable.

Intel provides a source-to-source translation tool from existing OpenACC implementation to OpenMP target one \cite{Intel_ACC2OMP} instead of providing OpenACC-supported compilers for Intel GPU.
AMD provides a similar tool named GPUFORT \cite{AMD_GPUFORT}.
Such source-to-source translators help developers migrate to the OpenMP target; however, if users keep using the original OpenACC codes, then the developers must maintain the two kinds of codes.
It detracts from the merit of directive-based programming, which allows for the coexistence of CPU and GPU implementation in a single code, minimizing maintenance costs.
Also, some notations in OpenACC and OpenMP target are not analogous, implying that maintaining the converted code requires knowledge of notation in OpenMP target.
\LibName{} reduces both the maintenance and learning costs because the adopted interfaces support both the OpenACC and OpenMP target, and OpenACC and OpenMP-like interfaces are also provided.
An alternative approach is Clacc \cite{Denny2018}, an OpenACC-supporting compiler built on the OpenMP support in Clang and LLVM.
The provided compiler helps OpenACC codes run on AMD and Intel GPUs; however, it is not guaranteed that all functions and benefits of compilers provided by GPU-manufacturing vendors will be exploited.

The contribution of this work, \LibName{}, realizes
\begin{itemize}
    \item unification of interfaces to exploit OpenACC and OpenMP target directives for GPU offloading,
    \item alleviation of vendor lock-in in the directive-based GPU programming,
    \item reduction of learning costs both for experienced and novice developers, and
    \item easy performance comparison of OpenACC and OpenMP target offloaded codes.
\end{itemize}
The remainder of the manuscript is organized as follows.
\S\ref{sec:implementation} shows the difficulty of a naive implementation, our solution, \LibName{}'s user interface with caution, and some implementation details.
\S\ref{sec:validation} validates \LibName{}'s features and the measured performance.
Finally, \S\ref{sec:summary} concludes this work.

\section{Implementation}
\label{sec:implementation}
\S\ref{subsec:outline} provides an example of a naive implementation compatible with the OpenACC and OpenMP target and points out the difficulty, followed by our solution adopted in \LibName{}.
The following subsections introduce \LibName{}'s user interface with precautions for users (\S\ref{subsec:interface}) and provide implementation details (\S\ref{subsec:implementation}).

\subsection{How to switch OpenACC and OpenMP target}
\label{subsec:outline}
Compilation enabling OpenACC defines the preprocessor macro \texttt{\_OPENACC}.
The most straightforward implementation of switching OpenACC and OpenMP target is like \texttt{\#ifdef \_OPENACC ... \#else ... \#endif}, which offloads the code to the GPU by OpenACC under the OpenACC-enabled compilation or by OpenMP target under the OpenACC-disabled compilation.
However, NVIDIA HPC SDK, which supports both OpenACC and OpenMP target, accepts the compile option as \texttt{-acc=multicore -mp=gpu}, indicating the parallelization for multicore CPUs by OpenACC and the GPU-offloading by OpenMP target directives.
In this case, the above implementation cannot realize the intended behavior; in other words, checking \texttt{\_OPENACC}'s presence only is insufficient.

\begin{figure}[t]
    \begin{lstlisting}[label=lst:normal, caption=Example using both of OpenACC and OpenMP target.]
#if defined(OFFLOAD_BY_OPENACC)
#if defined(OFFLOAD_BY_OPENACC_KERNELS)
<#red##pragma acc kernels
#pragma acc loop#>
#else
<#brown##pragma acc parallel
#pragma acc loop#>
#endif
#endif
#if defined(OFFLOAD_BY_OPENMP_TARGET)
#if defined(OFFLOAD_BY_OPENMP_TARGET_LOOP)
<#blue##pragma omp target teams loop#>
#else
<#green##pragma omp target teams distribute parallel for#>
#endif
#endif
  for (int i = 0; i < N_i; i++) {
    // loop body A
  }

#if defined(OFFLOAD_BY_OPENACC)
#if defined(OFFLOAD_BY_OPENACC_KERNELS)
<#red##pragma acc kernels
#pragma acc loop#>
#else
<#brown##pragma acc parallel
#pragma acc loop#>
#endif
#endif
#if defined(OFFLOAD_BY_OPENMP_TARGET)
#if defined(OFFLOAD_BY_OPENMP_TARGET_LOOP)
<#blue##pragma omp target teams loop#>
#else
<#green##pragma omp target teams distribute parallel for#>
#endif
#endif
  for (int j = 0; j < N_j; j++) {
    // loop body B
  }
  \end{lstlisting}
\end{figure}
The appropriate switching of OpenACC and OpenMP target requires an indication of which directive is in charge of the GPU offloading, such as putting \texttt{-DOFFLOAD\_BY\_OPENACC} at the compilation.
Furthermore, multiple methods in GPU offloading (\texttt{kernels}/\texttt{parallel} constructs in OpenACC, \texttt{loop}/\texttt{distribute} directives in OpenMP target) are available.
Such bifurcation leads to a cumbersome implementation, as shown in List.~\ref{lst:normal}.
Although the presented code works correctly, its poor visibility and multiple insertions of a bunch of directives hamper the adoption of this implementation style in GPU offloading of scientific codes.

\begin{figure}[t]
    \begin{lstlisting}[label=lst:macro, caption=Example using OpenACC or OpenMP target via macro.]
#if defined(OFFLOAD_BY_OPENACC)
#if defined(OFFLOAD_BY_OPENACC_KERNELS)
<#red##define OFFLOAD() _Pragma("acc kernels") _Pragma("acc loop")#>
#else
<#brown##define OFFLOAD() _Pragma("acc parallel") _Pragma("acc loop")#>
#endif
#endif
#if defined(OFFLOAD_BY_OPENMP_TARGET)
#if defined(OFFLOAD_BY_OPENMP_TARGET_LOOP)
<#blue##define OFFLOAD() _Pragma("omp target teams loop")#>
#else
<#green##define OFFLOAD() _Pragma("omp target teams distribute parallel for")#>
#endif
#endif

  OFFLOAD()
  for (int i = 0; i < N_i; i++) {
    // loop body A
  }

  OFFLOAD()
  for (int j = 0; j < N_j; j++) {
    // loop body B
  }
    \end{lstlisting}
\end{figure}
C/C++ provides \texttt{\_Pragma()}-style notation of directives instead of \texttt{\#pragma}-style, while the latter is widely adopted.
Using the former style notation, we can express directives as preprocessor macros.
List.~\ref{lst:macro} is a macro-based implementation of List.~\ref{lst:normal}, and it improves the visibility of the implementation by better-coordinating directives.

Extracting the definitions of preprocessor macros to external libraries further improves the visibility of the main codes.
Here, we have developed a library named \LibName{} containing the preprocessor macros corresponding to directives in the OpenACC and OpenMP target to facilitate easy implementation that is compatible with OpenACC and OpenMP target.
Actual GPU offloading requires dedicated controlling of each directive by putting optional clauses and lists of variables (e.g., operation and list of variables must be specified to \texttt{reduction} clause).
Since the possible users include experts who are already familiar with at least one of OpenACC and OpenMP target and beginners with no experience and knowledge of OpenACC and OpenMP target, \LibName{}'s user interface should minimize the learning cost for all users whose prerequisite knowledge is entirely different.

\subsection{User interface}
\label{subsec:interface}
This subsection introduces representative and frequently used directives and clauses (see \S\ref{app:API} for the rest of the APIs).
Then, guidance to hide or relax the differences between OpenACC and OpenMP target and instructions on how to switch backends of \LibName{} are presented.

\begin{sidewaystable*}
    \footnotesize
    \caption{Correspondence of representative directives.}
    \label{tab:directive}
    \centering
    \begin{tblr}{
        width={\linewidth},
        colspec={X[6]X[8]X[1]},
        cells={halign=l},
        hline{1,2,7,10,13,16,19,22,25,28,31,34,36}={solid},
        vline{2}={solid},
        column{1,2}={font={\ttfamily}},
        row{1} = {halign=c, font={\bfseries}},
        cell{2,7,10,13,16,19,22,25,28,31,34}{1}={font={\bfseries\ttfamily}}
    }
        input & output & backend\\
        OFFLOAD(...) & & \\
        PRAGMA\_ACC\_KERNELS\_LOOP(...) & \_Pragma("acc kernels \_\_VA\_ARGS\_\_") \_Pragma("acc loop \_\_VA\_ARGS\_\_") & OpenACC (kernels) \\
        PRAGMA\_ACC\_PARALLEL\_LOOP(...) & \_Pragma("acc parallel \_\_VA\_ARGS\_\_") \_Pragma("acc loop \_\_VA\_ARGS\_\_") & OpenACC (parallel) \\
        PRAGMA\_OMP\_TARGET\_TEAMS\_LOOP(...) & \_Pragma("omp target teams loop \_\_VA\_ARGS\_\_") & OpenMP (loop) \\
        PRAGMA\_OMP\_TARGET\_TEAMS\_DISTRIBUTE\_PARALLEL\_FOR(...) & \_Pragma("omp target teams distribute parallel for \_\_VA\_ARGS\_\_") & OpenMP (distribute) \\
        MALLOC\_ON\_DEVICE(...) & & \\
        PRAGMA\_ACC\_ENTER\_DATA\_CREATE(...) & \_Pragma("acc enter data create(\_\_VA\_ARGS\_\_)") & OpenACC \\
        PRAGMA\_OMP\_TARGET\_ENTER\_DATA\_MAP\_ALLOC(...) & \_Pragma("omp target enter data map(alloc: \_\_VA\_ARGS\_\_)") & OpenMP \\
        FREE\_FROM\_DEVICE(...) & & \\
        PRAGMA\_ACC\_EXIT\_DATA\_DELETE(...) & \_Pragma("acc exit data delete(\_\_VA\_ARGS\_\_)") & OpenACC \\
        PRAGMA\_OMP\_TARGET\_EXIT\_DATA\_MAP\_DELETE(...) & \_Pragma("omp target exit data map(delete: \_\_VA\_ARGS\_\_)") & OpenMP \\
        MEMCPY\_D2H(...) & & \\
        PRAGMA\_ACC\_UPDATE\_HOST(...) & \_Pragma("acc update host(\_\_VA\_ARGS\_\_)") & OpenACC \\
        PRAGMA\_OMP\_TARGET\_UPDATE\_FROM(...) & \_Pragma("omp target update from(\_\_VA\_ARGS\_\_)") & OpenMP \\
        MEMCPY\_H2D(...) & & \\
        PRAGMA\_ACC\_UPDATE\_DEVICE(...) & \_Pragma("acc update device(\_\_VA\_ARGS\_\_)") & OpenACC \\
        PRAGMA\_OMP\_TARGET\_UPDATE\_TO(...) & \_Pragma("omp target update to(\_\_VA\_ARGS\_\_)") & OpenMP \\
        DATA\_ACCESS\_BY\_DEVICE(...) & & \\
        PRAGMA\_ACC\_DATA(...) & \_Pragma("acc data \_\_VA\_ARGS\_\_") & OpenACC \\
        PRAGMA\_OMP\_TARGET\_DATA(...) & \_Pragma("omp target data \_\_VA\_ARGS\_\_") & OpenMP \\
        DATA\_ACCESS\_BY\_HOST(...) & & \\
        PRAGMA\_ACC\_HOST\_DATA(...) & \_Pragma("acc host\_data \_\_VA\_ARGS\_\_") & OpenACC \\
        PRAGMA\_OMP\_TARGET\_DATA(...) & \_Pragma("omp target data \_\_VA\_ARGS\_\_") & OpenMP \\
        SYNCHRONIZE(...) & & \\
        PRAGMA\_ACC\_WAIT(...) & \_Pragma("acc wait \_\_VA\_ARGS\_\_") & OpenACC \\
        PRAGMA\_OMP\_TARGET\_TASKWAIT(...) & \_Pragma("omp taskwait \_\_VA\_ARGS\_\_") & OpenMP \\
        ATOMIC(...) & & \\
        PRAGMA\_ACC\_ATOMIC(...) & \_Pragma("acc atomic \_\_VA\_ARGS\_\_") & OpenACC \\
        PRAGMA\_OMP\_TARGET\_ATOMIC(...) & \_Pragma("omp atomic \_\_VA\_ARGS\_\_") & OpenMP \\
        DECLARE\_OFFLOADED(...) & & \\
        PRAGMA\_ACC\_ROUTINE(...) & \_Pragma("acc routine \_\_VA\_ARGS\_\_") & OpenACC \\
        PRAGMA\_OMP\_DECLARE\_TARGET(...) & \_Pragma("omp declare target \_\_VA\_ARGS\_\_") & OpenMP \\
        DECLARE\_OFFLOADED\_END & & \\
        PRAGMA\_OMP\_END\_DECLARE\_TARGET & \_Pragma("omp end declare target") & OpenMP \\
    \end{tblr}
\end{sidewaystable*}
Tab.~\ref{tab:directive} indicates the relationship between input and output of representative directives in \LibName{} (see \S\ref{app:API} for the rest of the APIs).
\LibName{} provides three types of input interfaces:
(1) intuitive notations directly indicating the processing details like \texttt{OFFLOAD()} and \texttt{MEMCPY\_D2H()} (bold inputs in Tab.~\ref{tab:directive}),
(2) OpenACC-like notations as \texttt{PRAGMA\_ACC\_KERNELS\_LOOP()} and \texttt{PRAGMA\_ACC\_UPDATE\_HOST()}, and
(3) OpenMP-like notations as \texttt{PRAGMA\_OMP\_TARGET\_TEAMS\_LOOP()} and \texttt{PRAGMA\_OMP\_TARGET\_UPDATE\_FROM()}.
Users can combine these three notations because \LibName{} internally converts inputs to appropriate outputs.
The later section provides examples of the mixed implementation of the intuitive and OpenACC-like notations (see \S\ref{subsec:N-body} and \S\ref{subsec:diffusion3D}).
When users do not set the backend (OpenACC or OpenMP target) at the compilation, \LibName{} generates the thread-parallelized code using OpenMP for multicore CPUs, for example, by converting \texttt{OFFLOAD()} into \texttt{\_Pragma("omp parallel for")}.

\begin{table*}[t]
    \caption{Correspondence of representative clauses.}
    \label{tab:clause}
    \centering
    \begin{tblr}{
        width={\linewidth},
        cells={halign=l},
        hline{1,3,5,7,9,11,13}={solid},
        vline{3}={solid},
        column{1,2,3}={font={\ttfamily}},
        row{1,2}={halign=c, font={\bfseries}},
        cell{1}{1}={r=1,c=2}{halign=c, valign=m},
        cell{1}{3,4}={r=2,c=1}{halign=c, valign=m},
        cell{3,5,7,9,11}{1}={r=2,c=1}{halign=l, valign=m, font={\bfseries\ttfamily}},
        cell{7,9}{3}={r=2,c=1}{halign=l, valign=m}
    }
        input & input & output & backend\\
        intuitive notation & OpenACC/OpenMP-like notation & output & backend\\
        AS\_INDEPENDENT & ACC\_CLAUSE\_INDEPENDENT & independent & OpenACC\\
        AS\_INDEPENDENT & OMP\_TARGET\_CLAUSE\_SIMD & simd & OpenMP\\
        NUM\_THREADS(n) & ACC\_CLAUSE\_VECTOR\_LENGTH(n) & vector\_length(n) & OpenACC\\
        NUM\_THREADS(n) & OMP\_TARGET\_CLAUSE\_THREAD\_LIMIT(n) & thread\_limit(n) & OpenMP\\
        COLLAPSE(n) & ACC\_CLAUES\_COLLAPSE(n) & collapse(n) & OpenACC\\
        COLLAPSE(n) & OMP\_TARGET\_CLAUSE\_COLLAPSE(n) & collapse(n) & OpenMP\\
        REDUCTION(...) & ACC\_CLAUSE\_REDUCTION(...) & reduction(\_\_VA\_ARGS\_\_) & OpenACC\\
        REDUCTION(...) & OMP\_TARGET\_CLAUSE\_REDUCTION(...) & reduction(\_\_VA\_ARGS\_\_) & OpenMP\\
        AS\_ASYNC(...) & ACC\_CLAUSE\_ASYNC(...) & async(\_\_VA\_ARGS\_\_) & OpenACC\\
        AS\_ASYNC(...) & OMP\_TARGET\_CLAUSE\_NOWAIT & nowait & OpenMP\\
    \end{tblr}
\end{table*}
Directives are often accompanied by optional clauses and lists.
\LibName{} provides three types of notations for clauses (Tab.~\ref{tab:clause}), similar to directives.
In OpenMP-like notation, the indication of clauses related to OpenMP target directives as \texttt{PRAGMA\_OMP\_TARGET\_ATOMIC()} or \texttt{OMP\_TARGET\_CLAUSE\_COLLAPSE()} is required even for clauses available in OpenMP for multicore CPUs (\texttt{atomic} or \texttt{collapse}, respectively).
This indication helps \LibName{} in the OpenACC backend discriminate between directives/clauses converted to OpenACC (inputs with \texttt{TARGET}) and directives/clauses for multicore CPUs (inputs without \texttt{TARGET}).

The interface to indicate the optional clauses in \LibName{} differs from ordinary directives like OpenACC and OpenMP.
\LibName{} requests the comma-separated notation as \reviewtt{OFFLOAD(AS\_INDEPENDENT, ACC\_CLAUSE\_VECTOR\_LENGTH(128), OMP\_TARGET\_CLAUSE\_COLLAPSE(3))} for optional clauses.
This requirement comes from the necessity of filtering valid inputs and invalid (ignored) inputs for each OpenACC or OpenMP target directive (see \S\ref{subsec:implementation} for details).
As shown in the above example, users can combine three types of notations for clauses even in a single directive.

Because the implementation of \LibName{} is focused on providing preprocessor macros and converting them to appropriate directives and clauses in the specified backend, users should be aware of some cautions.
For example, indicating \texttt{independent} (OpenACC) or \texttt{simd} (OpenMP target) in the GPU offloading directives needs special care.
Adding the above options in the GPU offloading directives helps performance improvement in many cases by enhancing the vectorization.
The corresponding macros in \LibName{} are \texttt{AS\_INDEPENDENT}, \texttt{ACC\_CLAUSE\_INDEPENDENT}, and \texttt{OMP\_TARGET\_CLAUSE\_SIMD}.
When one specifies an option \texttt{AS\_INDEPENDENT} in the OpenMP target backend, \LibName{} regards the input as \texttt{simd} and generates a directive like \texttt{\_Pragma("omp target teams distribute parallel for simd")}.
However, in OpenMP, \texttt{simd} is not a clause and is a part of the combined construct.
It means that inserting another clause between \texttt{for} and \texttt{simd} causes an error in the above example.
Therefore, users must specify \texttt{AS\_INDEPENDENT} (, \texttt{ACC\_CLAUSE\_INDEPENDENT}, or \texttt{OMP\_TARGET\_CLAUSE\_SIMD}) at the head of all optional inputs.

There are further cautions regarding \LibName{}'s usage, which originates from the missing correspondence in the OpenACC and OpenMP target.
The \texttt{target data} directive in OpenMP target acts as the sum set of \texttt{data} and \texttt{host\_data} directives in OpenACC; the relation is not the injection.
Conversion from OpenACC to OpenMP target is quite simple: replace of \texttt{acc data} and \texttt{acc host\_data} to \texttt{omp target data}.
On the other hand, adequate splitting from \texttt{omp target data} to \texttt{acc data} or \texttt{acc host\_data} is mandatory for appropriate conversion from OpenMP target to OpenACC.
To realize the appropriate conversion,
\LibName{} recommends using (1) alternate macros \texttt{DATA\_ACCESS\_BY\_DEVICE(...)} and \texttt{PRAGMA\_HOST\_DATA(...)} or (2) OpenACC-like notation instead of using \texttt{PRAGMA\_OMP\_TARGET\_DATA(...)}.
The \texttt{routine} directive in OpenACC or \texttt{declare target} directive in OpenMP target indicates that the specified functions are GPU-offloaded.
In OpenACC, \texttt{routine} directive covers the function block, and an explicit indication of the end of the offloaded region is not required.
On the other hand, \texttt{declare target} directive in OpenMP target needs to add \texttt{end declare target} directive at the end of the offloaded region.
Thus, the appropriate insertion of \texttt{DECLARE\_OFFLOADED\_END} or \texttt{PRAGMA\_OMP\_END\_DECLARE\_TARGET} is necessary to run OpenACC-like implementation in the OpenMP target backend.
Since the added macros here are ignored by the OpenACC backend, this implementation will not affect OpenACC.

\begin{table*}[t]
    \caption{Compilation flags to switch backend directives.}
    \label{tab:switch}
    \centering
    \begin{tblr}{
        width={\linewidth},
        cells={halign=l},
        hline{1,2,6}={solid},
        column{1}={font={\ttfamily}},
        row{1}={font={\bfseries}}
    }
        compilation flag & backend\\
        -DOFFLOAD\_BY\_OPENACC & OpenACC (kernels)  \\
        -DOFFLOAD\_BY\_OPENACC -DOFFLOAD\_BY\_OPENACC\_PARALLEL & OpenACC (parallel)  \\
        -DOFFLOAD\_BY\_OPENMP\_TARGET & OpenMP target (loop)  \\
        -DOFFLOAD\_BY\_OPENMP\_TARGET -DOFFLOAD\_BY\_OPENMP\_TARGET\_DISTRIBUTE & OpenMP target (distribute)  \\
    \end{tblr}
\end{table*}
Users can select \LibName{}'s backend by specifying the compilation flags shown in Tab.~\ref{tab:switch}.
Indication of \texttt{-DOFFLOAD\_BY\_OPENACC} activates the OpenACC backend and offloads specified loops using \texttt{kernels} construct in default mode (except for \reviewtt{PRAGMA\_ACC\_PARALLEL\_LOOP(...)}, which uses \texttt{parallel} constructs).
Adding \texttt{-DOFFLOAD\_BY\_OPENACC\_PARALLEL} switches the default construct to \texttt{parallel} from \texttt{kernels}.
Similarly, \texttt{-DOFFLOAD\_BY\_OPENMP\_TARGET} activates the OpenMP target backend, exploits \texttt{loop} directives in the default, and adding \reviewtt{-DOFFLOAD\_BY\_OPENMP\_TARGET\_DISTRIBUTE} replaces the default directive to \texttt{distribute} instead of \texttt{loop}.
\LibName{} disables \texttt{OFFLOAD\_BY\_OPENMP\_TARGET} and exploits OpenACC as the backend, when both \texttt{-DOFFLOAD\_BY\_OPENACC} and \texttt{-DOFFLOAD\_BY\_OPENMP\_TARGET} are specified.
On the contrary, if neither \texttt{-DOFFLOAD\_BY\_OPENACC} nor \texttt{-DOFFLOAD\_BY\_OPENMP\_TARGET} is specified, \LibName{} sets OpenMP backend and performs thread-parallelization for multicore CPUs (e.g., \texttt{OFFLOAD()} is converted to \texttt{\_Pragma("omp parallel for")}).

\subsection{Internal implementation of \LibName{}}
\label{subsec:implementation}
Implementing macros corresponding to directives and clauses in OpenACC and OpenMP target is the main effort in the development of \LibName{}; however, additional efforts are required to accomplish the implementation of \LibName{}.
This subsection denotes the additional efforts and their necessity.
The design to simplify the implementation is also presented.

Converting OpenACC and OpenMP target each other is straightforward if the appropriate destination is registered.
However, a simple replacement of macro definitions is insufficient to switch OpenACC and OpenMP target.
For example, OpenMP target directive \texttt{\_Pragma("omp target teams loop collapse(3) thread\_limit(128)")} corresponds to \reviewtt{\_Pragma("acc kernels vector\_length(128)") \_Pragma("acc loop collapse(3)")} in OpenACC.
In this case, optional clauses in the corresponding specification \texttt{OFFLOAD(NUM\_THREADS(128), COLLAPSE(3))} must be distributed appropriately to \texttt{kernels} and \texttt{loop} constructs in OpenACC.
To realize this feature, \LibName{} internally sets flags to determine which clause is applicable in the directive, and then \LibName{} transforms only applicable clauses into OpenACC or OpenMP target clauses.
We have used publicly available code snippet\footnote{\url{https://github.com/pfultz2/Cloak}} \cite{macro_tips} to match two values inside the preprocessor macros to implement the filtering function.
The filtering function realizes the appropriate transformation to OpenACC by enabling applicable clauses after passing all optional clauses in \texttt{OFFLOAD(...)} to \texttt{kernels} and \texttt{loop} directives as candidates.

In the transformation from OpenACC to OpenMP target, two OpenACC clauses are sometimes gathered into a single OpenMP target directive.
It requires unifying multiple independent macros into a single macro; however, it is hard to implement using only preprocessor macros.
\LibName{} realizes this feature by providing macros like \texttt{PRAGMA\_ACC\_KERNELS\_LOOP(...)}, which makes it appear as a single directive (in fact, this macro is internally divided into two OpenACC directives, as shown in Tab.~\ref{tab:directive}).

Finally, we have simplified the implementation for OpenMP backend as follows.
Directives in OpenMP are usually combined ones; for example, \texttt{\_Pragma("omp target teams loop")} is a combined construct of \texttt{target} and \texttt{teams loop} (obviously, \texttt{teams loop} is a compound of \texttt{teams} and \texttt{loop}).
This construct accepts the sum set of clauses for \texttt{target}, \texttt{teams}, and \texttt{loop}.
We have simplified \LibName{}'s implementation by utilizing the above feature: \LibName{} equips flags for minimal directives or constructs like \texttt{teams} and \texttt{loop} but does not equip flags for combined directives or constructs like \texttt{teams loop}.
The determination process of the filtering function in \LibName{} uses the logical disjunction of individual evaluations for each clause to prevent the same clause from appearing multiple times, even if the specified clause is applicable both in \texttt{target} and \texttt{teams}.

\section{Validation and performance measurements}
\label{sec:validation}
This section validates the functionality of \LibName{} and measures the performance of $N$-body simulation (compute-intensive application, \S\ref{subsec:N-body}) and 3D diffusion equation (memory-intensive application, \S\ref{subsec:diffusion3D}).
We used \Hopper{}, \GraceHopper{}, \Aldebaran{}, and \PVC{} (hereafter, \HopperShort{}, \GraceHopperShort{}, \AldebaranShort{}, and \PVCshort{}) for the validation (see Tab.~\ref{tab:GPUspec} for the specifications of GPUs).
\begin{table*}[t]
    \caption{GPU specifications.}
    \label{tab:GPUspec}
    \centering
    \begin{talltblr}[
        label=none, entry=none,
        note{\textdagger}={GPU boost clock for FP32 non-Tensor Core operations for NVIDIA GPU, GPU boost clock for AMD and Intel GPUs.},
        note{\textdaggerdbl}={Thermal Design Power.},
        note{\textparagraph}={including CPU (NVIDIA Grace) and its memory.}
    ]{
        width={\linewidth},
        cells={halign=c},
        hline{1,2,12}={solid},
        vline{2}={solid},
        row{1}={font={\bfseries}},
        cell{4}{1}={r=2,c=1}{valign=m}
    }
        GPU & \Hopper{} & \GraceHopper{} & \Aldebaran{} & \PVC{} \\
        FP32 performance & \SI[round-mode=figures, round-precision=3]{66.908160}{\tera\flops} & \SI[round-mode=figures, round-precision=3]{66.908160}{\tera\flops} & \SI[round-mode=figures, round-precision=3]{22.630400}{\tera\flops} & \SI[round-mode=figures, round-precision=3]{22.2208}{\tera\flops}\\
        FP32 parallelism & \num{16896} & \num{16896} & \num{6656} & \num{7168} \\
        Number of units & \SI{132}{SMs} & \SI{132}{SMs} & \SI{104}{CUs} & \SI{448}{EUs} \\
        Number of units & 128 FP32 cores per SM & 128 FP32 cores per SM & 4x 16-wide SIMD per CU & 16-wide SIMD \\
        Clock frequency\TblrNote{\textdagger} & \SI{1980}{\mega\hertz} & \SI{1980}{\mega\hertz} & \SI{1700}{\mega\hertz} & \SI{1550}{\mega\hertz} \\
        Global memory & HBM3 \SI{80}{\giga\byte} & HBM3 \SI{96}{\giga\byte} & HBM2e \SI{64}{\giga\byte} & HBM2e \SI{48}{\giga\byte}\\
        Memory bandwidth & \SI{3.36}{\tera\bytes} & \SI[round-mode=figures, round-precision=3]{4.022}{\tera\bytes} & \SI{1.64}{\tera\bytes} & \SI{1.23}{\tera\bytes}\\
        L2 cache & \SI{50}{\mega\byte} & \SI{60}{\mega\byte} & \SI{16}{\mega\byte} & \SI{204}{\mega\byte}\\
        Shared memory & \SI{228}{K\byte} & \SI{228}{K\byte} & \SI{64}{K\byte} & \SI{64}{K\byte}\\
        TDP\TblrNote{\textdaggerdbl} & \SI{700}{\watt} & \SI{1000}{\watt}\TblrNote{\textparagraph} & \SI{300}{\watt} & \SI{300}{\watt}\\
    \end{talltblr}
\end{table*}
\begin{table*}[t]
    \caption{Measured environments.}
    \label{tab:environments}
    \centering
    \begin{talltblr}[
        label=none, entry=none,
        note{\textdagger}={We installed \acppFull{} using LLVM 18.1.7.}
    ]{
        width={\linewidth},
        cells={halign=c},
        hline{1,2,8}={solid},
        vline{2}={solid},
        row{1}={font={\bfseries}},
        cell{6}{1}={r=2,c=1}{valign=m}
    }
        GPU & \Hopper{} & \GraceHopper{} & \Aldebaran{} & \PVC{} \\
        CPU & Intel Xeon Platinum 8468 & NVIDIA Grace & AMD EPYC 7713 & Intel Xeon Platinum 8468 \\
        & $\SI{48}{cores} \times \SI{2}{sockets}$ & \SI{72}{cores} & $\SI{64}{cores} \times \SI{2}{sockets}$ & $\SI{48}{cores} \times \SI{2}{sockets}$\\
        & \SI{2.1}{\giga\hertz} & \SI{3.0}{\giga\hertz} & \SI{2.0}{\giga\hertz} & \SI{2.1}{\giga\hertz}\\
        OS & Rocky Linux 9.2 & Rocky Linux 9.4 & Rocky Linux 9.3 & Rocky Linux 9.3\\
        Compiler & CUDA 12.3.103 & CUDA 12.4.131 & ROCm 6.0.2 & \icpxFull{} 2024.1.0 \\
        Compiler & NVIDIA HPC SDK 24.3-0 & NVIDIA HPC SDK 24.5-1 & \acppFull{} 24.02.0\TblrNote{\textdagger} & \\
    \end{talltblr}
\end{table*}
\begin{table*}[t]
    \caption{Compile options in performance measurements.}
    \label{tab:options}
    \centering
    \begin{tblr}{
        width={\linewidth},
        colspec={X[1]X[9]},
        cells={halign=l},
        hline{1,2,6}={solid},
        column{1,2}={font={\ttfamily}},
        row{1}={font={\bfseries}}
    }
        Compiler & Compile option\\
        nvc++ & -Ofast -acc=gpu -gpu=cc90 \\
        nvc++ & -Ofast -mp=gpu -gpu=cc90 \\
        amdclang++ & -Ofast -target x86\_64-pc-linux-gnu -fopenmp -fopenmp-targets=amdgcn-amd-amdhsa -Xopenmp-target=amdgcn-amd-amdhsa -march=gfx90a\\
        icpx & -Ofast -fiopenmp -fopenmp-targets=spir64\_gen -Xs "-device pvc" \\
    \end{tblr}
\end{table*}
Tab.~\ref{tab:environments} lists the environments for the validation and performance measurements.
We used compilers provided by GPU vendors for OpenACC or OpenMP target, namely, NVIDIA HPC SDK for \HopperShort{} and \GraceHopperShort{}, AMD ROCm for \AldebaranShort{}, and \icpxFull{} for \PVCshort{}.
All the above compilers support OpenMP target directives, and only NVIDIA HPC SDK supports OpenACC.
Tab.~\ref{tab:options} summarizes the standard options for each compiler in the performance measurements.
The listed options include a standard performance optimizer \texttt{-Ofast} in recent compilers and the specification of the target GPU in each platform and compiler.

\subsection{$N$-body simulation}
\label{subsec:N-body}
\begin{figure}[t]
    \begin{lstlisting}[label=lst:nbody, caption=Implementation of $N$-body code using \LibName{}.]
void calc_acc(const int Ni, float4 *ipos, float4 *iacc, const int Nj, float4 *jpos, const float eps) {
  <#red#OFFLOAD(AS_INDEPENDENT, NUM_THREADS(NTHREADS))#>
  for (int i = 0; i < Ni; i++) {
    float4 pi = ipos[i];
    pi.w = eps * eps;
    float4 ai = {0.0F, 0.0F, 0.0F, 0.0F};
    <#blue#PRAGMA_ACC_LOOP(ACC_CLAUSE_SEQ)#>
    for (int j = 0; j < Nj; j++) {
      const float4 pj = jpos[j];
      float4 rji;
      rji.x = pj.x - pi.x;
      rji.y = pj.y - pi.y;
      rji.z = pj.z - pi.z;
      const auto r2 = std::fma(rji.z, rji.z, std::fma(rji.y, rji.y, std::fma(rji.x, rji.x, pi.w)));
      rji.w = 1.0F / std::sqrt(r2);
      rji.w *= rji.w * rji.w;
      rji.w *= pj.w;
      ai.x = std::fma(rji.x, rji.w, ai.x);
      ai.y = std::fma(rji.y, rji.w, ai.y);
      ai.z = std::fma(rji.z, rji.w, ai.z);
#ifdef CALCULATE_POTENTIAL
      ai.w = std::fma(r2, rji.w, ai.w);
#endif  // CALCULATE_POTENTIAL
    }
    iacc[i] = ai;
  }
}
  \end{lstlisting}
\end{figure}
$N$-body simulation, which calculates the time evolution of the gravitational many-body system by solving Newton's equation of motion, is a representative compute-intensive application frequently employed in astrophysical studies.
The fundamental equation of $N$-body simulations is
\begin{equation}
    \odv[order=2]{\bm{r}_i}{t} = \sum_{j = 0, j \neq i}^{N - 1} \frac{G m_j \pab{\bm{r}_j - \bm{r}_i}}{\pab{\vab{\bm{r}_j - \bm{r}_i}^2 + \epsilon^2}^{3/2}},
    \label{eq:acceleration}
\end{equation}
where $\bm{r}_i$ and $m_i$ are the position and mass of the $i$th particle out of $N$ particles, $G$ is the gravitational constant, and $\epsilon$ is the gravitational softening introduced to remove divergence owing to division by zero.
The implementation of the gravity calculation, based on the direct method, using \LibName{} is as shown in List.~\ref{lst:nbody}.
The macro \texttt{OFFLOAD(AS\_INDEPENDENT, NUM\_THREADS(NTHREADS))}, highlighted in red, corresponds to the directive for GPU offloading.
Because performance depends on the number of threads per block, we suggested that the compilers via \texttt{NTHREADS}--determined by a parameter survey on each environment--.
We also inserted the blue highlighted macro, \reviewtt{PRAGMA\_ACC\_LOOP(ACC\_CLAUSE\_SEQ)}, which is converted to \texttt{\_Pragma("acc loop seq")} in OpenACC so as not to parallelize the corresponding loop in OpenACC offloading.
The statement is ignored in OpenMP target offloading.
Without the statement in OpenACC implementation, the observed performance decreases owing to additional atomic operations required to parallelize the innermost loop.

The highlighted macros in List.~\ref{lst:nbody} complete the GPU offloading in both OpenACC and OpenMP target backends.
The number of lines necessary for GPU porting is similar to that of directive programming (OpenACC and OpenMP target) and much smaller than that of native languages (CUDA, HIP, and SYCL).
This single implementation can run on NVIDIA GPU (by OpenACC or OpenMP target), AMD/Intel GPUs (by OpenMP target), and multicore CPUs (by OpenMP); namely, there is no need to maintain multiple codes for multiple platforms or languages.

We added \texttt{-Mfprelaxed=rsqrt -gpu=fastmath} to the compilation option shown in Tab.~\ref{tab:options} on NVIDIA GPUs.
The added options are virtually mandatory in $N$-body simulations because removing them halves the performance (\texttt{-Ofast} cannot activate the fast, approximated arithmetic for the reciprocal square root).
The specified \texttt{fastmath} suggests the utilization of fast math libraries but seems to imply flushing subnormal numbers to zero, as the similar \texttt{--use\_fast\_math} option in CUDA implicitly adds \texttt{--ftz=true} option.
To keep a similar level of performance optimizations (at least by the compiler options), we added \texttt{-fgpu-flush-denormals-to-zero} for \AldebaranShort{} and \texttt{-ftz} for \PVCshort{}.

\begin{figure*}[t]
    \centering
    \includegraphics[width = \linewidth, pagebox = cropbox, clip]{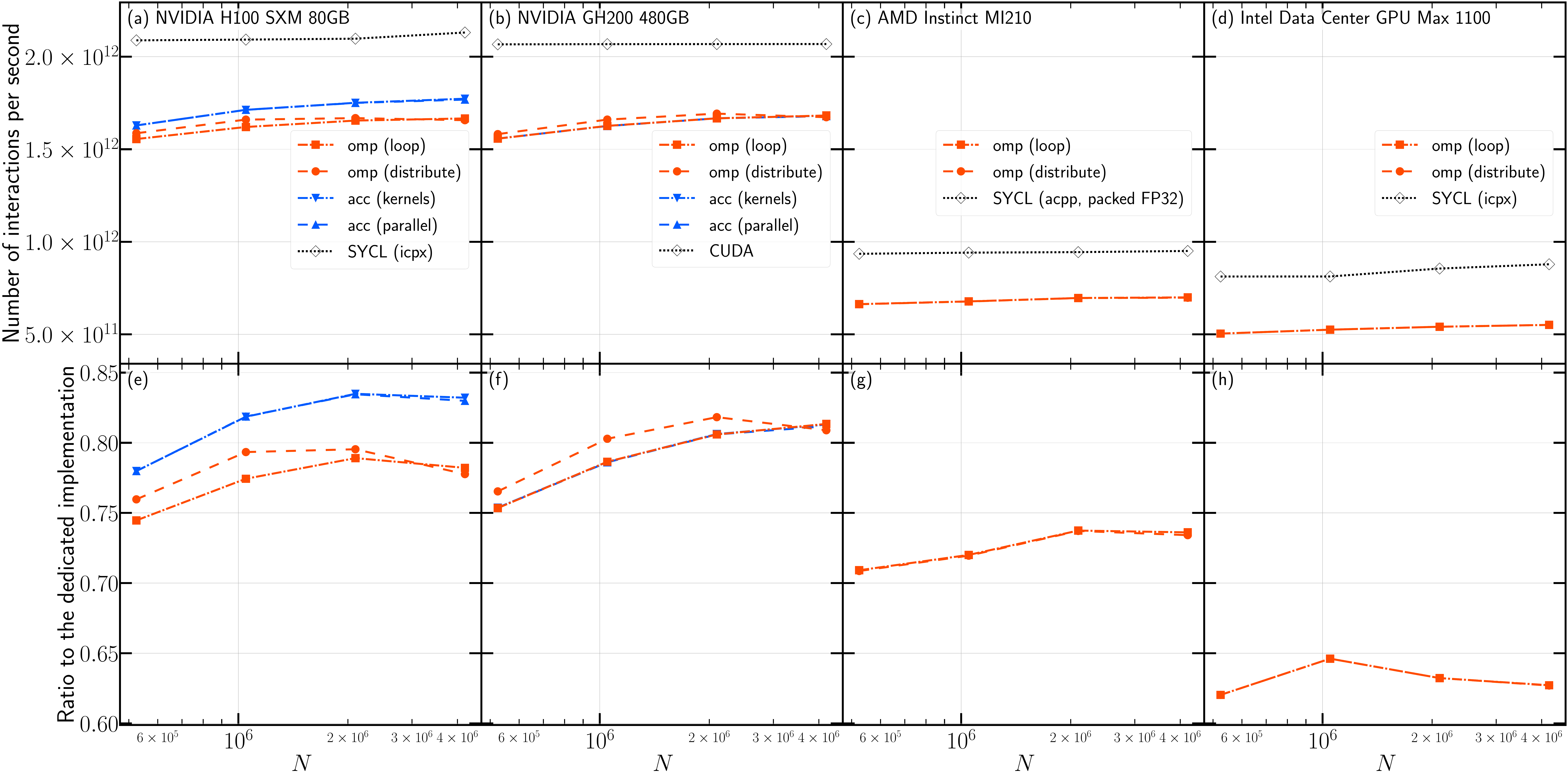}
    \caption{
        Measured performance of $N$-body simulations.
        The upper panels show the number of processed interaction pairs per second (the best performance in ten measurements) as a function of the number of $N$-body particles $N$.
        The open diamonds with a dotted line indicate the measured performance of the fastest implementation in each environment \cite{MikiHanawa2024}.
        The lower panels exhibit the performance ratio of OpenACC/OpenMP-offloaded implementations from the fastest implementation in \cite{MikiHanawa2024}.
        Each panel shows the measured performance on \Hopper{}, \GraceHopper{}, \Aldebaran{}, and \PVC{} from left to right.
        }
    \label{fig:nbody}
\end{figure*}
\LibName{} succeeded in offloading the kernels for OpenMP target (\texttt{loop} and \texttt{distribute} directives) on all four GPUs, and for OpenACC (\texttt{kernels} and \texttt{parallel} constructs) on NVIDIA GPUs.
Fig.~\ref{fig:nbody} shows the measured performance.
\LibName{} provides two backends for \texttt{OFFLOAD()} in OpenMP target (\texttt{loop} and \texttt{distribute} directives).
Their performances are almost identical on AMD and Intel GPUs, while \texttt{distribute} directive provides slightly higher performance on NVIDIA GPUs.
OpenACC backends are only available on NVIDIA GPUs, and the performance of \texttt{kernels} and \texttt{parallel} constructs is nearly indistinguishable.
OpenACC achieves superior performance than OpenMP target on \HopperShort{}, but the OpenACC performance on \GraceHopperShort{} is similar to that of OpenMP target \texttt{loop}.
The origin of the different behavior on NVIDIA GPUs is curious because the GPU part is the same, especially in $N$-body simulation.
The different versions of NVIDIA HPC SDK or graphics driver might alter the results.

Reference \cite{MikiHanawa2024} comprehensively investigated the fast and performance-portable implementation of $N$-body simulation on NVIDIA/AMD/Intel GPUs.
They found that (1) SYCL code compiled by \icpxFull{} was \num[round-mode=figures, round-precision=2]{1.6730840591406573}\% faster than CUDA on \HopperShort{}, (2) CUDA was the fastest on \GraceHopperShort{} owing to missing availability of \icpxFull{}, and (3) HIP C++ code and SYCL code compiled by \acppFull{} achieved almost identical performance, with the SYCL code achieving a slightly (\num[round-mode=figures, round-precision=2]{0.029845115549934333019613781688}\%) higher performance on \AldebaranShort{}.
Fig.~\ref{fig:nbody} shows the measured performance of their optimized implementation using CUDA/HIP/SYCL in the upper panels and indicates the performance ratio of the \LibName{}-offloaded performance in the lower panels.
Directive-based implementations lack some dedicated performance optimizations like exploitation of the shared memory, increase of instruction-level parallelism, and utilization of packed FP32 instructions (in AMD CDNA2 architecture); therefore, the observed performance of the \LibName{}-based offloaded performance is around \numrange[range-phrase=--]{60}{80}\% of the dedicated implementations.
The highest performances (the number of processed interaction pairs per second) in each environment are
$\SI[round-mode=figures, round-precision=3]{1.772994e12}{\second^{-1}}$ (\num[round-mode=figures, round-precision=3]{83.21544048245406121}\% of \cite{MikiHanawa2024}, OpenACC \texttt{kernels}) on \HopperShort{},
$\SI[round-mode=figures, round-precision=3]{1.682406e12}{\second^{-1}}$ (\num[round-mode=figures, round-precision=3]{81.34249448459821331}\% of \cite{MikiHanawa2024}, OpenMP target \texttt{loop}) on \GraceHopperShort{},
$\SI[round-mode=figures, round-precision=3]{6.995868e11}{\second^{-1}}$ (\num[round-mode=figures, round-precision=3]{73.36208658462527741}\% of \cite{MikiHanawa2024} with packed FP32 instructions and \num[round-mode=figures, round-precision=3]{97.53578087328516217}\% of \cite{MikiHanawa2024} with vector FP32 instructions, OpenMP target \texttt{loop}) on \AldebaranShort{}, and
$\SI[round-mode=figures, round-precision=3]{5.509536e11}{\second^{-1}}$ (\num[round-mode=figures, round-precision=3]{62.72471084717709907}\% of \cite{MikiHanawa2024}, OpenMP target \texttt{loop}) on \PVCshort{}.

\subsection{3D diffusion equation}
\label{subsec:diffusion3D}
\begin{figure*}
    \begin{lstlisting}[label=lst:diffusion, caption=Implementation of 3D diffusion equation using \LibName{}.]
#define INDEX(nx, ny, nz, ii, jj, kk) ((kk) + (nz) * ((jj) + (ny) * (ii)))
#define IMIN(a, b) (((a) < (b)) ? (a) : (b))
#define IMAX(a, b) (((a) > (b)) ? (a) : (b))

void diffusion3d(int nx, int ny, int nz, float dx, float dy, float dz, float dt, float kappa, const float *restrict f, float *restrict fn) {
  const float ce = kappa * dt / (dx * dx);  const float cw = ce;
  const float cn = kappa * dt / (dy * dy);  const float cs = cn;
  const float ct = kappa * dt / (dz * dz);  const float cb = ct;
  const float cc = 1.0F - (ce + cw + cn + cs + ct + cb);
  <#red#OFFLOAD(AS_INDEPENDENT, COLLAPSE(3), ACC_CLAUSE_PRESENT(f, fn))#>
  for (int i = 0; i < nx; i++) {
    for (int j = 0; j < ny; j++) {
      for (int k = 0; k < nz; k++) {
        const int ix = INDEX(nx, ny, nz, i, j, k);
        const int ip = INDEX(nx, ny, nz, IMIN(i + 1, nx - 1), j, k);
        const int im = INDEX(nx, ny, nz, IMAX(i - 1, 0), j, k);
        const int jp = INDEX(nx, ny, nz, i, IMIN(j + 1, ny - 1), k);
        const int jm = INDEX(nx, ny, nz, i, IMAX(j - 1, 0), k);
        const int kp = INDEX(nx, ny, nz, i, j, IMIN(k + 1, nz - 1));
        const int km = INDEX(nx, ny, nz, i, j, IMAX(k - 1, 0));
        fn[ix] = cc * f[ix] + ce * f[ip] + cw * f[im] + cn * f[jp] + cs * f[jm] + ct * f[kp] + cb * f[km];
      }
    }
  }
}
  \end{lstlisting}
\end{figure*}
The three-dimensional diffusion equation, which solves the time evolution of the distribution of diffusing materials, is a representative example of memory-intensive applications.
The equation is written as
\begin{equation}
    \pdv{u}{t} = a \nabla^2 u,
\end{equation}
where $u$ is the density of material and $a$ is the diffusion coeffiencient.
List.~\ref{lst:diffusion} shows the implementation using \LibName{}, and red-hilighted \reviewtt{OFFLOAD(AS\_INDEPENDENT, COLLAPSE(3), ACC\_CLAUSE\_PRESENT(f, fn))} is converted to directives for GPU offloading.
Like the $N$-body code shown in the previous subsection, the highlighted macro in List.~\ref{lst:diffusion} completes the GPU offloading in both OpenACC and OpenMP target backends with a similar effort in directive programming.

\begin{figure*}
    \centering
    \includegraphics[width = \linewidth, pagebox = cropbox, clip]{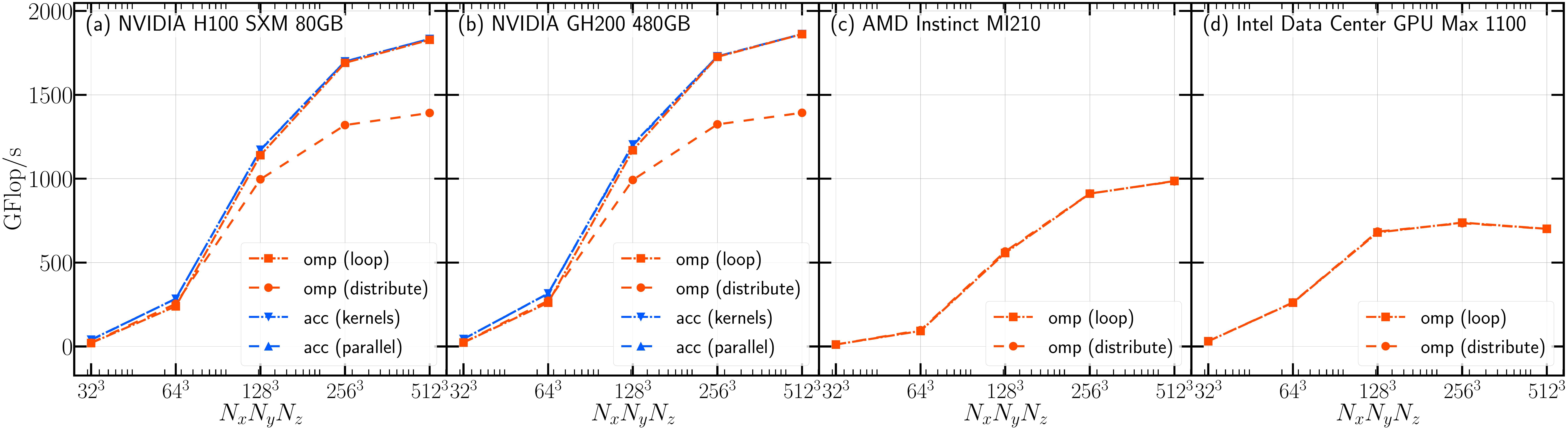}
    \caption{
        Measured performance of 3D diffusion equation.
        The panels display the best performance in ten measurements as a function of the total number of meshes $N_x N_y N_z$.
        The panels compare the measured performance of OpenMP target with \texttt{loop} (the red squares with a dot-dashed line), OpenMP target with \texttt{distribute} (the red circles with a dashed line), OpenACC with \texttt{kernels} (the blue lower triangles with a dot-dashed line), and OpenACC with \texttt{parallel} (the blue upper triangles with a dashed line).
        Each panel shows the measured performance on \Hopper{}, \GraceHopper{}, \Aldebaran{}, and \PVC{} from left to right.
        }
    \label{fig:diffusion}
\end{figure*}
\LibName{} succeeded in offloading the kernels for OpenMP target (\texttt{loop} and \texttt{distribute} directives) on all four GPUs and for OpenACC (\texttt{kernels} and \texttt{parallel} constructs) on NVIDIA GPUs.
Fig.~\ref{fig:diffusion} shows the measured performance.
On NVIDIA and Intel GPUs, performance without \texttt{-Ofast} option was higher than that with \texttt{-Ofast}; thus, the figure shows the performance without \texttt{-Ofast} option.
NVIDIA GPUs provide similar performance for OpenMP target with \texttt{loop} directive and OpenACC backends, while the \texttt{distribute} directive in OpenMP target achieves a substantially lower performance.
The performance of \texttt{loop} and \texttt{distribute} directives on AMD and Intel GPUs are almost indistinguishable.

The highest performance on \HopperShort{} and \GraceHopperShort{} is \SI[round-mode=figures, round-precision=3]{1.8332450000000001}{\tera\flops} (OpenACC \texttt{kernels}) and \SI[round-mode=figures, round-precision=3]{1.862265}{\tera\flops} (OpenACC \texttt{parallel}), respectively, which corresponds to \SI[round-mode=figures, round-precision=3]{4.512603076923077169}{\tera\bytes} and \SI[round-mode=figures, round-precision=3]{4.584036923076923077}{\tera\bytes} in memory bandwidth since the byte-per-flop ratio of 3D diffusion equation is \num[round-mode=figures, round-precision=3]{2.461538461538461538}.
The observed values exceed the global memory bandwidth; therefore, the cache's bandwidth would limit the performance.
It would be the reason for the weak or no contribution from the enhanced global-memory bandwidth of \GraceHopperShort{} over \HopperShort{} to the observed performance.
The \texttt{loop} directive delivers the best performance on \AldebaranShort{} and \PVCshort{}, with \SI[round-mode=figures, round-precision=3]{986.0937}{\giga\flops} (\SI[round-mode=figures, round-precision=3]{2.427307569230769231}{\tera\bytes}) and \SI[round-mode=figures, round-precision=3]{738.2486}{\giga\flops} (\SI[round-mode=figures, round-precision=3]{1.817227323076923077}{\tera\bytes}), respectively.
Although NVIDIA and AMD GPUs achieved the best performance at $N_x N_y N_z = 512^3 = \num{134217728}$, \PVCshort{} achieved the best at $N_x N_y N_z = 256^3 = \num{16777216}$.
The converted bandwidths also exceed the global memory bandwidth on AMD and Intel GPUs; thus, all GPU vendors' compilers generated well-optimized executables in the 3D diffusion equation.

\section{Conclusion}
\label{sec:summary}
Directive-based GPU offloading is a widely employed method for porting existing codes originally developed and optimized for multicore CPUs.
Both OpenACC and OpenMP target are a promising choice, but there are differences in available functions and applicable vendors.
We have developed a header-only library named \LibName{} to provide a unified interface supporting OpenACC and OpenMP target as backends.
\LibName{}'s three types of notations cover a wide range of users: intuitive notation for GPU beginners and OpenACC/OpenMP-like notations for experienced developers in directive-based GPU programming.
\S\ref{sec:validation} demonstrates \LibName{}'s ability in GPU offloading: NVIDIA/AMD/Intel GPUs returned appropriate results and achieved reasonable performance on the OpenMP target backend, and the OpenACC backend reproduced the results on NVIDIA GPUs where OpenACC was supported.
The unified implementation supporting OpenACC and OpenMP target eases fair performance comparisons between both methods, ensuring the same degree of performance optimization.

\appendices
\section{Available APIs in \LibName{}}
\label{app:API}
\LibName{} provides many APIs corresponding to OpenACC and OpenMP APIs.
Tables.~\ref{tab:directive} and \ref{tab:clause} show the representative and frequently used APIs only.
This section lists the rest of \LibName{}'s APIs.

\subsection{Directives for computation}
\label{app:API:compute}
Tables~\ref{tab:directive:comp:atomic}--\ref{tab:directive:comp:OpenMP} list directives for computation.
Bold texts indicating the intuitive notation are followed by the OpenACC or OpenMP correspondings.
Abstraction APIs for kernel launch are tabulated in Tab.~\ref{tab:directive:comp:abstraction}.
The tables for OpenACC/OpenMP-like APIs, the counterparts in OpenMP/OpenACC backends, and fallback mode are also provided.
If the direct counterpart is missing, then the input is automatically ignored by \LibName{}.

\begin{table*}[t]
    \caption{Directives for computation}
    \label{tab:directive:comp:atomic}
    \centering
    \begin{tblr}{
        width={\linewidth},
        cells={halign=l},
        hline{1,2,5,8,11,14}={solid},
        vline{2}={solid},
        column{1,2}={font={\ttfamily}},
        row{1}={halign=c, font={\bfseries}},
        cell{2,5,8,11}{1}={font={\bfseries\ttfamily}}
    }
    input & output & backend \\
    ATOMIC\_UPDATE & & \\
    PRAGMA\_ACC\_ATOMIC\_UPDATE         & \_Pragma("acc atomic update") & OpenACC \\
    PRAGMA\_OMP\_TARGET\_ATOMIC\_UPDATE & \_Pragma("omp atomic update") & OpenMP \\
    ATOMIC\_READ & & \\
    PRAGMA\_ACC\_ATOMIC\_READ & \_Pragma("acc atomic read")& OpenACC\\
    PRAGMA\_OMP\_TARGET\_ATOMIC\_READ & \_Pragma("omp atomic read") & OpenMP\\
    ATOMIC\_WRITE & & \\
    PRAGMA\_ACC\_ATOMIC\_WRITE & \_Pragma("acc atomic write") & OpenACC\\
    PRAGMA\_OMP\_TARGET\_ATOMIC\_WRITE & \_Pragma("omp atomic write") & OpenMP\\
    ATOMIC\_CAPTURE & & \\
    PRAGMA\_ACC\_ATOMIC\_CAPTURE & \_Pragma("acc atomic capture") & OpenACC\\
    PRAGMA\_OMP\_TARGET\_ATOMIC\_CAPTURE & \_Pragma("omp atomic capture") & OpenMP\\
    \end{tblr}
\end{table*}

\begin{sidewaystable*}[t]
    \footnotesize
    \caption{Abstraction API}
    \label{tab:directive:comp:abstraction}
    \centering
    \begin{tblr}{
        width={\linewidth},
        cells={halign=l},
        hline{1,2,5,8}={solid},
        vline{2}={solid},
        column{1,2}={font={\ttfamily}},
        row{1}={halign=c, font={\bfseries}}
    }
    input & output & backend \\
    PRAGMA\_ACC\_LAUNCH\_DEFAULT(...) & PRAGMA\_ACC\_KERNELS(\_\_VA\_ARGS\_\_) & OpenACC (kernels) \\
    PRAGMA\_ACC\_LAUNCH\_DEFAULT(...) & PRAGMA\_ACC\_PARALLEL(\_\_VA\_ARGS\_\_) & OpenACC (parallel) \\
    PRAGMA\_OMP\_TARGET\_LAUNCH\_DEFAULT(...) & PRAGMA\_OMP\_TARGET\_TEAMS(\_\_VA\_ARGS\_\_) & OpenMP \\
    PRAGMA\_ACC\_OFFLOADING\_DEFAULT(...) & PRAGMA\_ACC\_LAUNCH\_DEFAULT(\_\_VA\_ARGS\_\_) PRAGMA\_ACC\_LOOP(\_\_VA\_ARGS\_\_) & OpenACC \\
    PRAGMA\_OMP\_TARGET\_OFFLOADING\_DEFAULT(...) & PRAGMA\_OMP\_TARGET\_TEAMS\_LOOP(\_\_VA\_ARGS\_\_) & OpenMP (loop) \\
    PRAGMA\_OMP\_TARGET\_OFFLOADING\_DEFAULT(...) & PRAGMA\_OMP\_TARGET\_TEAMS\_DISTRIBUTE\_PARALLEL\_FOR(\_\_VA\_ARGS\_\_) & OpenMP (distribute)\\
    \end{tblr}

    \footnotesize
    \caption{OpenACC directives}
    \label{tab:directive:comp:OpenACC}
    \centering
    \begin{tblr}{
        width={\linewidth},
        cells={halign=l},
        hline{1,2,11}={solid},
        vline{2}={solid},
        column{1,2}={font={\ttfamily}},
        row{1}={halign=c, font={\bfseries}}
    }
    input & output & counterpart in OpenMP backend \\
    PRAGMA\_ACC\_PARALLEL(...) & \_Pragma("acc parallel \_\_VA\_ARGS\_\_") & \texttt{PRAGMA\_OMP\_TARGET\_OFFLOADING\_DEFAULT(\_\_VA\_ARGS\_\_)} \\
    PRAGMA\_ACC\_KERNELS(...)  & \_Pragma("acc kernels \_\_VA\_ARGS\_\_")  & \texttt{PRAGMA\_OMP\_TARGET\_OFFLOADING\_DEFAULT(\_\_VA\_ARGS\_\_)} \\
    PRAGMA\_ACC\_SERIAL(...)   & \_Pragma("acc serial \_\_VA\_ARGS\_\_")   & N/A (disregarded in OpenMP backend) \\
    PRAGMA\_ACC\_LOOP(...)     & \_Pragma("acc loop \_\_VA\_ARGS\_\_")     & N/A (disregarded in OpenMP backend) \\
    PRAGMA\_ACC\_CACHE(...)    & \_Pragma("acc cache(\_\_VA\_ARGS\_\_)")   & N/A (disregarded in OpenMP backend) \\
    PRAGMA\_ACC\_ATOMIC(...)   & \_Pragma("acc atomic \_\_VA\_ARGS\_\_")   & \texttt{PRAGMA\_OMP\_TARGET\_ATOMIC(\_\_VA\_ARGS\_\_)} \\
    PRAGMA\_ACC\_WAIT(...)     & \_Pragma("acc wait \_\_VA\_ARGS\_\_")     & \texttt{PRAGMA\_OMP\_TARGET\_TASKWAIT(\_\_VA\_ARGS\_\_)} \\
    PRAGMA\_ACC\_ROUTINE(...)  & \_Pragma("acc routine \_\_VA\_ARGS\_\_")  & \texttt{PRAGMA\_OMP\_DECLARE\_TARGET(\_\_VA\_ARGS\_\_)} \\
    PRAGMA\_ACC\_DECLARE(...)  & \_Pragma("acc declare \_\_VA\_ARGS\_\_")  & N/A (disregarded in OpenMP backend) \\
    \end{tblr}
\end{sidewaystable*}

\begin{sidewaystable*}[t]
    \footnotesize
    \caption{OpenMP target directives}
    \label{tab:directive:comp:OpenMP.target}
    \centering
    \begin{tblr}{
        width={\linewidth},
        colspec={X[1]X[1]X[1]X[1]},
        cells={halign=l},
        hline{1,2,19}={solid},
        vline{2}={solid},
        row{1}={halign=c, font={\bfseries}}
    }
    input & output & counterpart in OpenACC backend & counterpart in fallback mode (CPU execution without offloading) \\
    \reviewtt{PRAGMA\_OMP\_TARGET(...)} & \reviewtt{\_Pragma("omp target \_\_VA\_ARGS\_\_")} & \reviewtt{PRAGMA\_ACC(\_\_VA\_ARGS\_\_)} & N/A (disregarded in fallback mode) \\
    \reviewtt{PRAGMA\_OMP\_TARGET\_PARALLEL(...)} & \reviewtt{\_Pragma("omp target parallel \_\_VA\_ARGS\_\_")} & \reviewtt{PRAGMA\_ACC\_LAUNCH\_DEFAULT(\_\_VA\_ARGS\_\_)} & \reviewtt{PRAGMA\_OMP\_PARALLEL(\_\_VA\_ARGS\_\_)} \\
    \reviewtt{PRAGMA\_OMP\_TARGET\_PARALLEL\_FOR(...)} & \reviewtt{\_Pragma("omp target parallel for \_\_VA\_ARGS\_\_")} & \reviewtt{PRAGMA\_ACC\_OFFLOADING\_DEFAULT(\_\_VA\_ARGS\_\_)} & \reviewtt{PRAGMA\_OMP\_PARALLEL\_FOR(\_\_VA\_ARGS\_\_)} \\
    \reviewtt{PRAGMA\_OMP\_TARGET\_PARALLEL\_FOR\_SIMD(...)} & \reviewtt{\_Pragma("omp target parallel for simd \_\_VA\_ARGS\_\_")} & \reviewtt{PRAGMA\_ACC\_OFFLOADING\_DEFAULT(ACC\_CLAUSE\_INDEPENDENT, \#\#\_\_VA\_ARGS\_\_)} & \reviewtt{PRAGMA\_OMP\_PARALLEL\_FOR\_SIMD(\_\_VA\_ARGS\_\_)} \\
    \reviewtt{PRAGMA\_OMP\_TARGET\_PARALLEL\_LOOP(...)} & \reviewtt{\_Pragma("omp target parallel loop \_\_VA\_ARGS\_\_")} & \reviewtt{PRAGMA\_ACC\_OFFLOADING\_DEFAULT(\_\_VA\_ARGS\_\_)} & \reviewtt{PRAGMA\_OMP\_PARALLEL\_LOOP(\_\_VA\_ARGS\_\_)} \\
    \reviewtt{PRAGMA\_OMP\_TARGET\_SIMD(...)} & \reviewtt{\_Pragma("omp target simd \_\_VA\_ARGS\_\_")} & \reviewtt{PRAGMA\_ACC\_LAUNCH\_DEFAULT(ACC\_CLAUSE\_INDEPENDENT, \#\#\_\_VA\_ARGS\_\_)} & \reviewtt{PRAGMA\_OMP\_SIMD(\_\_VA\_ARGS\_\_)} \\
    \reviewtt{PRAGMA\_OMP\_TARGET\_TEAMS(...)} & \reviewtt{\_Pragma("omp target teams \_\_VA\_ARGS\_\_")} & \reviewtt{PRAGMA\_ACC\_LAUNCH\_DEFAULT(\_\_VA\_ARGS\_\_)} & \reviewtt{PRAGMA\_OMP\_TEAMS(\_\_VA\_ARGS\_\_)} \\
    \reviewtt{PRAGMA\_OMP\_TARGET\_TEAMS\_DISTRIBUTE(...)} & \reviewtt{\_Pragma("omp target teams distribute \_\_VA\_ARGS\_\_")} & \reviewtt{PRAGMA\_ACC\_LAUNCH\_DEFAULT(\_\_VA\_ARGS\_\_)} & \reviewtt{PRAGMA\_OMP\_TEAMS\_DISTRIBUTE(\_\_VA\_ARGS\_\_)} \\
    \reviewtt{PRAGMA\_OMP\_TARGET\_TEAMS\_DISTRIBUTE\_SIMD(...)} & \reviewtt{\_Pragma("omp target teams distribute simd \_\_VA\_ARGS\_\_")} & \reviewtt{PRAGMA\_ACC\_LAUNCH\_DEFAULT(ACC\_CLAUSE\_INDEPENDENT, \#\#\_\_VA\_ARGS\_\_)} & \reviewtt{PRAGMA\_OMP\_TEAMS\_DISTRIBUTE\_SIMD(\_\_VA\_ARGS\_\_)} \\
    \reviewtt{PRAGMA\_OMP\_TARGET\_TEAMS\_LOOP(...)} & \reviewtt{\_Pragma("omp target teams loop \_\_VA\_ARGS\_\_")} & \reviewtt{PRAGMA\_ACC\_OFFLOADING\_DEFAULT(\_\_VA\_ARGS\_\_)} & \reviewtt{PRAGMA\_OMP\_TEAMS\_LOOP(\_\_VA\_ARGS\_\_)} \\
    \reviewtt{PRAGMA\_OMP\_TARGET\_TEAMS\_DISTRIBUTE\_PARALLEL\_FOR(...)} & \reviewtt{\_Pragma("omp target teams distribute parallel for \_\_VA\_ARGS\_\_")} & \reviewtt{PRAGMA\_ACC\_OFFLOADING\_DEFAULT(\_\_VA\_ARGS\_\_)} & \reviewtt{PRAGMA\_OMP\_TEAMS\_DISTRIBUTE\_PARALLEL\_FOR(\_\_VA\_ARGS\_\_)} \\
    \reviewtt{PRAGMA\_OMP\_TARGET\_TEAMS\_DISTRIBUTE\_PARALLEL\_FOR\_SIMD(...)} & \reviewtt{\_Pragma("omp target teams distribute parallel for simd \_\_VA\_ARGS\_\_")} & \reviewtt{PRAGMA\_ACC\_OFFLOADING\_DEFAULT(ACC\_CLAUSE\_INDEPENDENT, \#\#\_\_VA\_ARGS\_\_)} & \reviewtt{PRAGMA\_OMP\_TEAMS\_DISTRIBUTE\_PARALLEL\_FOR\_SIMD(\_\_VA\_ARGS\_\_)} \\
    \reviewtt{PRAGMA\_OMP\_TARGET\_ATOMIC(...)} & \reviewtt{PRAGMA\_OMP\_ATOMIC(\_\_VA\_ARGS\_\_)} & \reviewtt{PRAGMA\_ACC\_ATOMIC(\_\_VA\_ARGS\_\_)} & \reviewtt{PRAGMA\_OMP\_ATOMIC(\_\_VA\_ARGS\_\_)} \\
    \reviewtt{PRAGMA\_OMP\_TARGET\_TASKWAIT(...)} & \reviewtt{PRAGMA\_OMP\_TASKWAIT(\_\_VA\_ARGS\_\_)} & \reviewtt{PRAGMA\_ACC\_WAIT(\_\_VA\_ARGS\_\_)} & \reviewtt{PRAGMA\_OMP\_TASKWAIT(\_\_VA\_ARGS\_\_)} \\
    \reviewtt{PRAGMA\_OMP\_DECLARE\_TARGET(...)} & \reviewtt{\_Pragma("omp declare target \_\_VA\_ARGS\_\_")} & \reviewtt{PRAGMA\_ACC\_ROUTINE(\_\_VA\_ARGS\_\_)} & N/A (disregarded in fallback mode) \\
    \reviewtt{PRAGMA\_OMP\_BEGIN\_DECLARE\_TARGET(...)} & \reviewtt{\_Pragma("omp begin declare target \_\_VA\_ARGS\_\_")} & \reviewtt{PRAGMA\_ACC\_ROUTINE(\_\_VA\_ARGS\_\_)} & N/A (disregarded in fallback mode) \\
    \reviewtt{PRAGMA\_OMP\_END\_DECLARE\_TARGET} & \reviewtt{\_Pragma("omp end declare target")} & N/A (disregarded in OpenACC backend) & N/A (disregarded in fallback mode) \\
    \end{tblr}
\end{sidewaystable*}

\begin{table*}[t]
    \caption{OpenMP directives}
    \label{tab:directive:comp:OpenMP}
    \centering
    \begin{tblr}{
        width={\linewidth},
        colspec={X[16]X[17]},
        cells={halign=l},
        hline{1,2,49}={solid},
        vline{2}={solid},
        column{1,2}={font={\ttfamily}},
        row{1}={halign=c, font={\bfseries}}
    }
    input & output \\
    PRAGMA\_OMP\_THREADPRIVATE(...) & \_Pragma("omp threadprivate(\_\_VA\_ARGS\_\_)") \\
    PRAGMA\_OMP\_SCAN(...) & \_Pragma("omp scan \_\_VA\_ARGS\_\_") \\
    PRAGMA\_OMP\_DECLARE\_SIMD(...) & \_Pragma("omp declare simd \_\_VA\_ARGS\_\_") \\
    PRAGMA\_OMP\_TILE(...) & \_Pragma("omp tile \_\_VA\_ARGS\_\_") \\
    PRAGMA\_OMP\_UNROLL(...) & \_Pragma("omp unroll \_\_VA\_ARGS\_\_") \\
    PRAGMA\_OMP\_PARALLEL(...) & \_Pragma("omp parallel \_\_VA\_ARGS\_\_") \\
    PRAGMA\_OMP\_TEAMS(...) & \_Pragma("omp teams \_\_VA\_ARGS\_\_") \\
    PRAGMA\_OMP\_SIMD(...) & \_Pragma("omp simd \_\_VA\_ARGS\_\_") \\
    PRAGMA\_OMP\_MASKED(...) & \_Pragma("omp masked \_\_VA\_ARGS\_\_") \\
    PRAGMA\_OMP\_SINGLE(...) & \_Pragma("omp single \_\_VA\_ARGS\_\_") \\
    PRAGMA\_OMP\_WORKSHARE(...) & \_Pragma("omp workshare \_\_VA\_ARGS\_\_") \\
    PRAGMA\_OMP\_SCOPE(...) & \_Pragma("omp scope \_\_VA\_ARGS\_\_") \\
    PRAGMA\_OMP\_SECTIONS(...) & \_Pragma("omp sections \_\_VA\_ARGS\_\_") \\
    PRAGMA\_OMP\_SECTION & \_Pragma("omp section") \\
    PRAGMA\_OMP\_FOR(...) & \_Pragma("omp for \_\_VA\_ARGS\_\_") \\
    PRAGMA\_OMP\_DISTRIBUTE(...) & \_Pragma("omp distribute \_\_VA\_ARGS\_\_") \\
    PRAGMA\_OMP\_LOOP(...) & \_Pragma("omp loop \_\_VA\_ARGS\_\_") \\
    PRAGMA\_OMP\_TASK(...) & \_Pragma("omp task \_\_VA\_ARGS\_\_") \\
    PRAGMA\_OMP\_TASKLOOP(...) & \_Pragma("omp taskloop \_\_VA\_ARGS\_\_") \\
    PRAGMA\_OMP\_TASKYIELD & \_Pragma("omp taskyield") \\
    PRAGMA\_OMP\_INTEROP(...) & \_Pragma("omp interop \_\_VA\_ARGS\_\_") \\
    PRAGMA\_OMP\_CRITICAL(...) & \_Pragma("omp critical \_\_VA\_ARGS\_\_") \\
    PRAGMA\_OMP\_BARRIER & \_Pragma("omp barrier") \\
    PRAGMA\_OMP\_TASKGROUP(...) & \_Pragma("omp taskgroup \_\_VA\_ARGS\_\_") \\
    PRAGMA\_OMP\_TASKWAIT(...) & \_Pragma("omp taskwait \_\_VA\_ARGS\_\_") \\
    PRAGMA\_OMP\_FLUSH(...) & \_Pragma("omp flush \_\_VA\_ARGS\_\_") \\
    PRAGMA\_OMP\_DEPOBJ(...) & \_Pragma("omp depobj \_\_VA\_ARGS\_\_") \\
    PRAGMA\_OMP\_ATOMIC(...) & \_Pragma("omp atomic \_\_VA\_ARGS\_\_") \\
    PRAGMA\_OMP\_ORDERED(...) & \_Pragma("omp ordered \_\_VA\_ARGS\_\_") \\
    PRAGMA\_OMP\_FOR\_SIMD(...) & \_Pragma("omp for simd \_\_VA\_ARGS\_\_") \\
    PRAGMA\_OMP\_DISTRIBUTE\_SIMD(...) & \_Pragma("omp distribute simd \_\_VA\_ARGS\_\_") \\
    PRAGMA\_OMP\_DISTRIBUTE\_PARALLEL\_FOR(...) & \_Pragma("omp distribute parallel for \_\_VA\_ARGS\_\_") \\
    PRAGMA\_OMP\_DISTRIBUTE\_PARALLEL\_FOR\_SIMD(...) & \_Pragma("omp distribute parallel for simd \_\_VA\_ARGS\_\_") \\
    PRAGMA\_OMP\_TASKLOOP\_SIMD(...) & \_Pragma("omp taskloop simd \_\_VA\_ARGS\_\_") \\
    PRAGMA\_OMP\_PARALLEL\_FOR(...) & \_Pragma("omp parallel for \_\_VA\_ARGS\_\_") \\
    PRAGMA\_OMP\_PARALLEL\_LOOP(...) & \_Pragma("omp parallel loop \_\_VA\_ARGS\_\_") \\
    PRAGMA\_OMP\_PARALLEL\_SECTIONS(...) & \_Pragma("omp parallel sections \_\_VA\_ARGS\_\_") \\
    PRAGMA\_OMP\_PARALLEL\_FOR\_SIMD(...) & \_Pragma("omp parallel for simd \_\_VA\_ARGS\_\_") \\
    PRAGMA\_OMP\_MASKED\_TASKLOOP(...) & \_Pragma("omp masked taskloop \_\_VA\_ARGS\_\_") \\
    PRAGMA\_OMP\_MASKED\_TASKLOOP\_SIMD(...) & \_Pragma("omp masked taskloop simd \_\_VA\_ARGS\_\_") \\
    PRAGMA\_OMP\_PARALLEL\_MASKED\_TASKLOOP(...) & \_Pragma("omp parallel masked taskloop \_\_VA\_ARGS\_\_") \\
    PRAGMA\_OMP\_PARALLEL\_MASKED\_TASKLOOP\_SIMD(...) & \_Pragma("omp parallel masked taskloop simd \_\_VA\_ARGS\_\_") \\
    PRAGMA\_OMP\_TEAMS\_DISTRIBUTE(...) & \_Pragma("omp teams distribute \_\_VA\_ARGS\_\_") \\
    PRAGMA\_OMP\_TEAMS\_DISTRIBUTE\_SIMD(...) & \_Pragma("omp teams distribute simd \_\_VA\_ARGS\_\_") \\
    PRAGMA\_OMP\_TEAMS\_DISTRIBUTE\_PARALLEL\_FOR(...) & \_Pragma("omp teams distribute parallel for \_\_VA\_ARGS\_\_") \\
    PRAGMA\_OMP\_TEAMS\_DISTRIBUTE\_PARALLEL\_FOR\_SIMD(...) & \_Pragma("omp teams distribute parallel for simd \_\_VA\_ARGS\_\_") \\
    PRAGMA\_OMP\_TEAMS\_LOOP(...) & \_Pragma("omp teams loop \_\_VA\_ARGS\_\_") \\
    \end{tblr}
\end{table*}

\subsection{Directives for memory and/or data transfer}
Tables~\ref{tab:directive:data}--\ref{tab:directive:data:OpenMP} exhibit the directives for memory manipulation and/or data transfer between host and device.

\begin{table*}[t]
    \caption{Directives for memory manipulation and/or data transfer}
    \label{tab:directive:data}
    \centering
    \begin{tblr}{
        width={\linewidth},
        colspec={X[5]X[6]X[1]},
        cells={halign=l},
        hline{1,2,12,15,17}={solid},
        vline{2}={solid},
        column{1,2}={font={\ttfamily}},
        row{1}={halign=c, font={\bfseries}},
        cell{12,15}{1}={font={\bfseries\ttfamily}}
    }
    input & output & backend \\
    PRAGMA\_ACC\_ENTER\_DATA(...) & \_Pragma("acc enter data \_\_VA\_ARGS\_\_") & OpenACC \\
    PRAGMA\_OMP\_TARGET\_ENTER\_DATA(...) & \_Pragma("omp target enter data \_\_VA\_ARGS\_\_") & OpenMP \\
    PRAGMA\_ACC\_ENTER\_DATA\_COPYIN(...) & \_Pragma("acc enter data copyin(\_\_VA\_ARGS\_\_)") & OpenACC \\
    PRAGMA\_OMP\_TARGET\_ENTER\_DATA\_MAP\_TO(...) & \_Pragma("omp target enter data map(to: \_\_VA\_ARGS\_\_)") & OpenMP \\
    PRAGMA\_ACC\_EXIT\_DATA(...) & \_Pragma("acc exit data \_\_VA\_ARGS\_\_") & OpenACC \\
    PRAGMA\_OMP\_TARGET\_EXIT\_DATA(...) & \_Pragma("omp target exit data \_\_VA\_ARGS\_\_") & OpenMP \\
    PRAGMA\_ACC\_EXIT\_DATA\_COPYOUT(...) & \_Pragma("acc exit data copyout(\_\_VA\_ARGS\_\_)") & OpenACC \\
    PRAGMA\_OMP\_TARGET\_EXIT\_DATA\_MAP\_FROM(...) & \_Pragma("omp target exit data map(from: \_\_VA\_ARGS\_\_)") & OpenMP \\
    PRAGMA\_ACC\_UPDATE(...) & \_Pragma("acc update \_\_VA\_ARGS\_\_") & OpenACC \\
    PRAGMA\_OMP\_TARGET\_UPDATE(...) & \_Pragma("omp target update \_\_VA\_ARGS\_\_") & OpenMP \\
    USE\_DEVICE\_DATA\_FROM\_HOST(...) & & \\
    PRAGMA\_ACC\_HOST\_DATA\_USE\_DEVICE(...) & \_Pragma("acc host\_data use\_device(\_\_VA\_ARGS\_\_)") & OpenACC \\
    PRAGMA\_OMP\_TARGET\_DATA\_USE\_DEVICE\_PTR(...) & \_Pragma("omp target data use\_device\_ptr(\_\_VA\_ARGS\_\_)") & OpenMP \\
    DECLARE\_DATA\_ON\_DEVICE(...) & & \\
    PRAGMA\_ACC\_DATA\_PRESENT(...) & \_Pragma("acc data present(\_\_VA\_ARGS\_\_)") & OpenACC (only) \\
    \end{tblr}
\end{table*}

\begin{table*}[t]
    \caption{OpenACC directives}
    \label{tab:directive:data:OpenACC}
    \centering
    \begin{tblr}{
        width={\linewidth},
        cells={halign=l},
        hline{1,2,7}={solid},
        vline{2}={solid},
        column{1,2,3}={font={\ttfamily}},
        row{1}={halign=c, font={\bfseries}}
    }
    input & output & counterpart in OpenMP backend \\
    PRAGMA\_ACC\_DATA(...) & \_Pragma("acc data \_\_VA\_ARGS\_\_") & PRAGMA\_OMP\_TARGET\_DATA(\_\_VA\_ARGS\_\_) \\
    PRAGMA\_ACC\_ENTER\_DATA(...) & \_Pragma("acc enter data \_\_VA\_ARGS\_\_") & PRAGMA\_OMP\_TARGET\_ENTER\_DATA(\_\_VA\_ARGS\_\_) \\
    PRAGMA\_ACC\_EXIT\_DATA(...) & \_Pragma("acc exit data \_\_VA\_ARGS\_\_") & PRAGMA\_OMP\_TARGET\_EXIT\_DATA(\_\_VA\_ARGS\_\_) \\
    PRAGMA\_ACC\_HOST\_DATA(...) & \_Pragma("acc host\_data \_\_VA\_ARGS\_\_") & PRAGMA\_OMP\_TARGET\_DATA(\_\_VA\_ARGS\_\_) \\
    PRAGMA\_ACC\_UPDATE(...) & \_Pragma("acc update \_\_VA\_ARGS\_\_") & PRAGMA\_OMP\_TARGET\_UPDATE(\_\_VA\_ARGS\_\_) \\
    \end{tblr}
\end{table*}

\begin{sidewaystable*}[t]
    \footnotesize
    \caption{OpenMP target directives}
    \label{tab:directive:data:OpenMP}
    \centering
    \begin{tblr}{
        width={\linewidth},
        colspec={X[2]X[3]X[2]X[1]},
        cells={halign=l},
        hline{1,2,6}={solid},
        vline{2}={solid},
        column{1,2,3}={font={\ttfamily}},
        row{1}={halign=c, font={\bfseries}}
    }
    input & output & counterpart in OpenACC backend & output in fallback mode (CPU execution without offloading)\\
    PRAGMA\_OMP\_TARGET\_DATA(...) & \_Pragma("omp target data \_\_VA\_ARGS\_\_") & PRAGMA\_ACC\_DATA(\_\_VA\_ARGS\_\_) & N/A (disregarded in fallback mode) \\
    PRAGMA\_OMP\_TARGET\_ENTER\_DATA(...) & \_Pragma("omp target enter data \_\_VA\_ARGS\_\_") & PRAGMA\_ACC\_ENTER\_DATA(\_\_VA\_ARGS\_\_) & N/A (disregarded in fallback mode) \\
    PRAGMA\_OMP\_TARGET\_EXIT\_DATA(...) & \_Pragma("omp target exit data \_\_VA\_ARGS\_\_") & PRAGMA\_ACC\_EXIT\_DATA(\_\_VA\_ARGS\_\_) & N/A (disregarded in fallback mode) \\
    PRAGMA\_OMP\_TARGET\_UPDATE(...) & \_Pragma("omp target update \_\_VA\_ARGS\_\_") & PRAGMA\_ACC\_UPDATE(\_\_VA\_ARGS\_\_) & N/A (disregarded in fallback mode) \\
    \end{tblr}
\end{sidewaystable*}

\subsection{Available clauses}
\label{app:API:clause}
Tables~\ref{tab:clause:intuitive}--\ref{tab:clause:OpenMP.2} are lists of available clauses in \LibName{}.
Notation is shared with the tables in \S\ref{app:API:compute}.

\begin{table*}[t]
    \caption{Intuitive clauses}
    \label{tab:clause:intuitive}
    \centering
    \begin{tblr}{
        width={\linewidth},
        cells={halign=l},
        hline{1,2,4,7,9,11,13,15,18,21,24,27,30,33,36}={solid},
        vline{2}={solid},
        column{1,2}={font={\ttfamily}},
        row{1}={halign=c, font={\bfseries}},
        cell{2,4,7,9,11,13,15,18,21,24,27,30,33}{1}={font={\bfseries\ttfamily}}
    }
    input & output & backend \\
    AS\_SEQUENTIAL & & \\
    ACC\_CLAUSE\_SEQ & seq & OpenACC (only) \\
    NUM\_BLOCKS(n) & & \\
    ACC\_CLAUSE\_NUM\_WORKERS(n) & num\_workers(n) & OpenACC \\
    OMP\_TARGET\_CLAUSE\_NUM\_TEAMS(n) & num\_teams(n) & OpenMP \\
    NUM\_GRIDS(n) & & \\
    ACC\_CLAUSE\_NUM\_GANGS(n) & num\_gang(n) & OpenACC (only) \\
    AS\_THREAD & & \\
    ACC\_CLAUSE\_VECTOR & vector & OpenACC (only) \\
    AS\_BLOCK & & \\
    ACC\_CLAUSE\_WORKER & worker & OpenACC (only) \\
    AS\_GRID & & \\
    ACC\_CLAUSE\_GANG & gang & OpenACC (only) \\
    ENABLE\_IF(condition) & & \\
    ACC\_CLAUSE\_IF(condition) & if(condition) & OpenACC \\
    OMP\_TARGET\_CLAUSE\_IF(condition) & if(condition) & OpenMP \\
    AS\_PRIVATE(...) & & \\
    ACC\_CLAUSE\_PRIVATE(...) & private(\_\_VA\_ARGS\_\_) & OpenACC \\
    OMP\_TARGET\_CLAUSE\_PRIVATE(...) & private(\_\_VA\_ARGS\_\_) & OpenMP \\
    AS\_FIRSTPRIVATE(...) & & \\
    ACC\_CLAUSE\_FIRSTPRIVATE(...) & firstprivate(\_\_VA\_ARGS\_\_) & OpenACC \\
    OMP\_TARGET\_CLAUSE\_FIRSTPRIVATE(...) & firstprivate(\_\_VA\_ARGS\_\_) & OpenMP \\
    AS\_DEVICE\_PTR(...) & & \\
    ACC\_CLAUSE\_DEVICEPTR(...) & deviceptr(\_\_VA\_ARGS\_\_) & OpenACC \\
    OMP\_TARGET\_CLAUSE\_IS\_DEVICE\_PTR(...) & is\_device\_ptr(\_\_VA\_ARGS\_\_) & OpenMP \\
    COPY\_BEFORE\_AND\_AFTER\_EXEC(...) & & \\
    ACC\_CLAUSE\_COPY(...) & copy(\_\_VA\_ARGS\_\_) & OpenACC \\
    OMP\_TARGET\_CLAUSE\_MAP\_TOFROM(...) & map(tofrom: \_\_VA\_ARGS\_\_) & OpenMP \\
    COPY\_H2D\_BEFORE\_EXEC(...) & & \\
    ACC\_CLAUSE\_COPYIN(...) & copyin(\_\_VA\_ARGS\_\_) & OpenACC \\
    OMP\_TARGET\_CLAUSE\_MAP\_TO(...) & map(to: \_\_VA\_ARGS\_\_) & OpenMP \\
    COPY\_D2H\_AFTER\_EXEC(...) & & \\
    ACC\_CLAUSE\_COPYOUT(...) & copyout(\_\_VA\_ARGS\_\_) & OpenACC \\
    OMP\_TARGET\_CLAUSE\_MAP\_FROM(...) & map(from: \_\_VA\_ARGS\_\_) & OpenMP \\
    \end{tblr}
\end{table*}

\begin{table*}[t]
    \caption{OpenACC clauses}
    \label{tab:clause:OpenACC}
    \centering
    \begin{talltblr}[
        label=none, entry=none,
        note{\textdagger}={notation as \texttt{gang(n)} is N/A},
        note{\textdaggerdbl}={notation as \texttt{worker(n)} is N/A},
        note{\textparagraph}={notation as \texttt{vector(n)} is N/A}
    ]{
        width={\linewidth},
        cells={halign=l},
        hline{1,2,48}={solid},
        vline{2}={solid},
        column{1,2,3}={font={\ttfamily}},
        row{1}={halign=c, font={\bfseries}}
    }
    input & output & counterpart in OpenMP backend \\
    ACC\_CLAUSE\_IF(condition) & if(condition) & OMP\_TARGET\_CLAUSE\_IF(condition) \\
    ACC\_CLAUSE\_SELF(...) & self(\_\_VA\_ARGS\_\_) & \textrm{N/A (disregarded in OpenMP backend)} \\
    ACC\_CLAUSE\_DEFAULT(mode) & default(mode) & \textrm{N/A (disregarded in OpenMP backend)} \\
    ACC\_CLAUSE\_DEFAULT\_NONE & default(none) & OMP\_TARGET\_CLAUSE\_DEFAULTMAP\_NONE \\
    ACC\_CLAUSE\_DEFAULT\_PRESENT & default(present) & OMP\_TARGET\_CLAUSE\_DEFAULTMAP\_PRESENT \\
    ACC\_CLAUSE\_DEVICE\_TYPE(...) & device\_type(\_\_VA\_ARGS\_\_) & OMP\_TARGET\_CLAUSE\_DEVICE\_TYPE(\_\_VA\_ARGS\_\_) \\
    ACC\_CLAUSE\_ASYNC(...) & async(\_\_VA\_ARGS\_\_) & OMP\_TARGET\_CLAUSE\_NOWAIT \\
    ACC\_CLAUSE\_WAIT(...) & wait(\_\_VA\_ARGS\_\_) & OMP\_TARGET\_CLAUSE\_DEPEND\_IN(\_\_VA\_ARGS\_\_) \\
    ACC\_CLAUSE\_FINALIZE & finalize & \textrm{N/A (disregarded in OpenMP backend)} \\
    ACC\_CLAUSE\_NUM\_GANGS(n) & num\_gangs(n) & \textrm{N/A (disregarded in OpenMP backend)} \\
    ACC\_CLAUSE\_NUM\_WORKERS(n) & num\_workers(n) & OMP\_TARGET\_CLAUSE\_NUM\_TEAMS(n) \\
    ACC\_CLAUSE\_VECTOR\_LENGTH(n) & vector\_length(n) & OMP\_TARGET\_CLAUSE\_THREAD\_LIMIT(n) \\
    ACC\_CLAUSE\_REDUCTION(...) & reduction(\_\_VA\_ARGS\_\_) & OMP\_TARGET\_CLAUSE\_REDUCTION(\_\_VA\_ARGS\_\_) \\
    ACC\_CLAUSE\_PRIVATE(...) & private(\_\_VA\_ARGS\_\_) & OMP\_TARGET\_CLAUSE\_PRIVATE(\_\_VA\_ARGS\_\_) \\
    ACC\_CLAUSE\_FIRSTPRIVATE(...) & firstprivate(\_\_VA\_ARGS\_\_) & OMP\_TARGET\_CLAUSE\_FIRSTPRIVATE(\_\_VA\_ARGS\_\_) \\
    ACC\_CLAUSE\_COPY(...) & copy(\_\_VA\_ARGS\_\_) & OMP\_TARGET\_CLAUSE\_MAP\_TOFROM(\_\_VA\_ARGS\_\_) \\
    ACC\_CLAUSE\_COPYIN(...) & copyin(\_\_VA\_ARGS\_\_) & OMP\_TARGET\_CLAUSE\_MAP\_TO(\_\_VA\_ARGS\_\_) \\
    ACC\_CLAUSE\_COPYOUT(...) & copyout(\_\_VA\_ARGS\_\_) & OMP\_TARGET\_CLAUSE\_MAP\_FROM(\_\_VA\_ARGS\_\_) \\
    ACC\_CLAUSE\_CREATE(...) & create(\_\_VA\_ARGS\_\_) & OMP\_TARGET\_CLAUSE\_MAP\_ALLOC(\_\_VA\_ARGS\_\_) \\
    ACC\_CLAUSE\_NO\_CREATE(...) & no\_create(\_\_VA\_ARGS\_\_) & \textrm{N/A (disregarded in OpenMP backend)} \\
    ACC\_CLAUSE\_DELETE(...) & delete(\_\_VA\_ARGS\_\_) & OMP\_TARGET\_CLAUSE\_MAP\_DELETE(\_\_VA\_ARGS\_\_) \\
    ACC\_CLAUSE\_PRESENT(...) & present(\_\_VA\_ARGS\_\_) & \textrm{N/A (disregarded in OpenMP backend)} \\
    ACC\_CLAUSE\_DEVICEPTR(...) & deviceptr(\_\_VA\_ARGS\_\_) & OMP\_TARGET\_CLAUSE\_IS\_DEVICE\_PTR(\_\_VA\_ARGS\_\_) \\
    ACC\_CLAUSE\_ATTACH(...) & attach(\_\_VA\_ARGS\_\_) & \textrm{N/A (disregarded in OpenMP backend)} \\
    ACC\_CLAUSE\_DETACH(...) & detach(\_\_VA\_ARGS\_\_) & \textrm{N/A (disregarded in OpenMP backend)} \\
    ACC\_CLAUSE\_USE\_DEVICE(...) & use\_device(\_\_VA\_ARGS\_\_) & OMP\_TARGET\_CLAUSE\_USE\_DEVICE\_PTR(\_\_VA\_ARGS\_\_) \\
    ACC\_CLAUSE\_IF\_PRESENT & if\_present & \textrm{N/A (disregarded in OpenMP backend)} \\
    ACC\_CLAUSE\_COLLAPSE(n) & collapse(n) & OMP\_TARGET\_CLAUSE\_COLLAPSE(n) \\
    ACC\_CLAUSE\_SEQ & seq & \textrm{N/A (disregarded in OpenMP backend)} \\
    ACC\_CLAUSE\_AUTO & auto & \textrm{N/A (disregarded in OpenMP backend)} \\
    ACC\_CLAUSE\_INDEPENDENT & independent & OMP\_TARGET\_CLAUSE\_SIMD \\
    ACC\_CLAUSE\_TILE(...) & tile(\_\_VA\_ARGS\_\_) & \textrm{N/A (disregarded in OpenMP backend)} \\
    ACC\_CLAUSE\_GANG & gang\TblrNote{\textdagger} & \textrm{N/A (disregarded in OpenMP backend)} \\
    ACC\_CLAUSE\_WORKER & worker\TblrNote{\textdaggerdbl} & \textrm{N/A (disregarded in OpenMP backend)} \\
    ACC\_CLAUSE\_VECTOR & vector\TblrNote{\textparagraph} & \textrm{N/A (disregarded in OpenMP backend)} \\
    ACC\_CLAUSE\_READ & read & OMP\_TARGET\_CLAUSE\_READ \\
    ACC\_CLAUSE\_WRITE & write & OMP\_TARGET\_CLAUSE\_WRITE \\
    ACC\_CLAUSE\_UPDATE & update & OMP\_TARGET\_CLAUSE\_UPDATE \\
    ACC\_CLAUSE\_CAPTURE & capture & OMP\_TARGET\_CLAUSE\_CAPTURE \\
    ACC\_CLAUSE\_HOST(...) & host(\_\_VA\_ARGS\_\_) & OMP\_TARGET\_CLAUSE\_FROM(\_\_VA\_ARGS\_\_) \\
    ACC\_CLAUSE\_DEVICE(...) & device(\_\_VA\_ARGS\_\_) & OMP\_TARGET\_CLAUSE\_TO(\_\_VA\_ARGS\_\_) \\
    ACC\_PASS\_LIST(...) & (\_\_VA\_ARGS\_\_) & OMP\_TARGET\_PASS\_LIST(\_\_VA\_ARGS\_\_) & \\
    ACC\_CLAUSE\_BIND(arg) & bind(arg) & N/A (disregarded in OpenMP backend) \\
    ACC\_CLAUSE\_NOHOST & nohost & OMP\_TARGET\_CLAUSE\_DEVICE\_TYPE(nohost) \\
    ACC\_CLAUSE\_DEVICE\_RESIDENT(...) & device\_resident(\_\_VA\_ARGS\_\_) & \textrm{N/A (disregarded in OpenMP backend)} \\
    ACC\_CLAUSE\_LINK(...) & link(\_\_VA\_ARGS\_\_) & \textrm{N/A (disregarded in OpenMP backend)} \\
    \end{talltblr}
\end{table*}

\begin{sidewaystable*}[t]
    \footnotesize
    \caption{OpenMP target clauses (1/3)}
    \label{tab:clause:OpenMP.target.1}
    \centering
    \begin{tblr}{
        width={\linewidth},
        colspec={XXXX},
        rowhead=2,
        cells={halign=l},
        hline{1,2,24}={solid},
        vline{2}={solid},
        row{1}={halign=c, font={\bfseries}}
    }
    input & output & counterpart in OpenACC backend & counterpart in fallback mode (CPU execution without offloading) \\
    \reviewtt{OMP\_TARGET\_CLAUSE\_ALIGNED(...)} & \reviewtt{OMP\_CLAUSE\_ALIGNED(\_\_VA\_ARGS\_\_)} & N/A (disregarded in OpenACC backend) & \reviewtt{OMP\_CLAUSE\_ALIGNED(\_\_VA\_ARGS\_\_)} \\
    \reviewtt{OMP\_TARGET\_CLAUSE\_SIMDLEN(len)} & \reviewtt{OMP\_CLAUSE\_SIMDLEN(len)} & N/A (disregarded in OpenACC backend) & \reviewtt{OMP\_CLAUSE\_SIMDLEN(len)} \\
    \reviewtt{OMP\_TARGET\_CLAUSE\_DEVICE\_TYPE(type)} & \reviewtt{device\_type(type)} & \reviewtt{ACC\_CLAUSE\_DEVICE\_TYPE(type)} &  N/A (disregarded in fallback mode) \\
    \reviewtt{OMP\_TARGET\_CLAUSE\_ENTER(...)} & \reviewtt{enter(\_\_VA\_ARGS\_\_)} & N/A (disregarded in OpenACC backend) & N/A (disregarded in fallback mode) \\
    \reviewtt{OMP\_TARGET\_CLAUSE\_INDIRECT(...)} & \reviewtt{indirect(\_\_VA\_ARGS\_\_)} &  N/A (disregarded in OpenACC backend) & N/A (disregarded in fallback mode) \\
    \reviewtt{OMP\_TARGET\_CLAUSE\_LINK(...)} & \reviewtt{link(\_\_VA\_ARGS\_\_)} &  N/A (disregarded in OpenACC backend) & N/A (disregarded in fallback mode) \\
    \reviewtt{OMP\_TARGET\_CLAUSE\_COPYIN(...)} & \reviewtt{OMP\_CLAUSE\_COPYIN(\_\_VA\_ARGS\_\_)} & \reviewtt{ACC\_CLAUSE\_COPYIN(\_\_VA\_ARGS\_\_)} & \reviewtt{OMP\_CLAUSE\_COPYIN(\_\_VA\_ARGS\_\_)} \\
    \reviewtt{OMP\_TARGET\_CLAUSE\_NUM\_THREADS(num)} & \reviewtt{OMP\_CLAUSE\_NUM\_THREADS(num)} & \reviewtt{ACC\_CLAUSE\_VECTOR\_LENGTH(num)} & \reviewtt{OMP\_CLAUSE\_NUM\_THREADS(num)} \\
    \reviewtt{OMP\_TARGET\_CLAUSE\_PROC\_BIND(attr)} & \reviewtt{OMP\_CLAUSE\_PROC\_BIND(attr)} & N/A (disregarded in OpenACC backend) & \reviewtt{OMP\_CLAUSE\_PROC\_BIND(attr)} \\
    \reviewtt{OMP\_TARGET\_CLAUSE\_NUM\_TEAMS(...)} & \reviewtt{OMP\_CLAUSE\_NUM\_TEAMS(\_\_VA\_ARGS\_\_)} & \reviewtt{ACC\_CLAUSE\_NUM\_WORKERS(\_\_VA\_ARGS\_\_)} & \reviewtt{OMP\_CLAUSE\_NUM\_TEAMS(\_\_VA\_ARGS\_\_)} \\
    \reviewtt{OMP\_TARGET\_CLAUSE\_THREAD\_LIMIT(num)} & \reviewtt{OMP\_CLAUSE\_THREAD\_LIMIT(num)} & \reviewtt{ACC\_CLAUSE\_VECTOR\_LENGTH(num)} & \reviewtt{OMP\_CLAUSE\_THREAD\_LIMIT(num)} \\
    \reviewtt{OMP\_TARGET\_CLAUSE\_NONTEMPORAL(...)} & \reviewtt{OMP\_CLAUSE\_NONTEMPORAL(\_\_VA\_ARGS\_\_)} & N/A (disregarded in OpenACC backend) & \reviewtt{OMP\_CLAUSE\_NONTEMPORAL(\_\_VA\_ARGS\_\_)} \\
    \reviewtt{OMP\_TARGET\_CLAUSE\_SAFELEN(len)} & \reviewtt{OMP\_CLAUSE\_SAFELEN(len)} & N/A (disregarded in OpenACC backend) & \reviewtt{OMP\_CLAUSE\_SAFELEN(len)} \\
    \reviewtt{OMP\_TARGET\_CLAUSE\_ORDERED(...)} & \reviewtt{OMP\_CLAUSE\_ORDERED(\_\_VA\_ARGS\_\_)} & N/A (disregarded in OpenACC backend) & \reviewtt{OMP\_CLAUSE\_ORDERED(\_\_VA\_ARGS\_\_)} \\
    \reviewtt{OMP\_TARGET\_CLAUSE\_SCHEDULE(...)} & \reviewtt{OMP\_CLAUSE\_SCHEDULE(\_\_VA\_ARGS\_\_)} &  N/A (disregarded in OpenACC backend) & \reviewtt{OMP\_CLAUSE\_SCHEDULE(\_\_VA\_ARGS\_\_)} \\
    \reviewtt{OMP\_TARGET\_CLAUSE\_DIST\_SCHEDULE(...)} & \reviewtt{OMP\_CLAUSE\_DIST\_SCHEDULE(\_\_VA\_ARGS\_\_)} &  N/A (disregarded in OpenACC backend) & \reviewtt{OMP\_CLAUSE\_DIST\_SCHEDULE(\_\_VA\_ARGS\_\_)} \\
    \reviewtt{OMP\_TARGET\_CLAUSE\_BIND(bind)} & \reviewtt{OMP\_CLAUSE\_BIND(bind)} & N/A (disregarded in OpenACC backend) & \reviewtt{OMP\_CLAUSE\_BIND(bind)} \\
    \reviewtt{OMP\_TARGET\_CLAUSE\_USE\_DEVICE\_PTR(...)} & \reviewtt{use\_device\_ptr(\_\_VA\_ARGS\_\_)} & \reviewtt{ACC\_CLAUSE\_USE\_DEVICE(\_\_VA\_ARGS\_\_)} & N/A (disregarded in fallback mode) \\
    \reviewtt{OMP\_TARGET\_CLAUSE\_USE\_DEVICE\_ADDR(...)} & \reviewtt{use\_device\_addr(\_\_VA\_ARGS\_\_)} & \reviewtt{ACC\_CLAUSE\_USE\_DEVICE(\_\_VA\_ARGS\_\_)} & N/A (disregarded in fallback mode) \\
    \reviewtt{OMP\_TARGET\_CLAUSE\_DEFAULTMAP(...)} & \reviewtt{defaultmap(\_\_VA\_ARGS\_\_)} & N/A (disregarded in OpenACC backend) & N/A (disregarded in fallback mode) \\
    \reviewtt{OMP\_TARGET\_CLAUSE\_DEFAULTMAP\_NONE} & \reviewtt{OMP\_TARGET\_CLAUSE\_DEFAULTMAP(none)} & \reviewtt{ACC\_CLAUSE\_DEFAULT\_NONE} & N/A (disregarded in fallback mode) \\
    \reviewtt{OMP\_TARGET\_CLAUSE\_DEFAULTMAP\_PRESENT} & \reviewtt{OMP\_TARGET\_CLAUSE\_DEFAULTMAP(present)} & \reviewtt{ACC\_CLAUSE\_DEFAULT\_PRESENT} & N/A (disregarded in fallback mode) \\
    \end{tblr}
\end{sidewaystable*}
\begin{sidewaystable*}[t]
    \footnotesize
    \caption{OpenMP target clauses (2/3)}
    \label{tab:clause:OpenMP.target.2}
    \centering
    \begin{tblr}{
        width={\linewidth},
        colspec={XXXX},
        rowhead=2,
        cells={halign=l},
        hline{1,2,26}={solid},
        vline{2}={solid},
        row{1}={halign=c, font={\bfseries}}
    }
    input & output & counterpart in OpenACC backend & counterpart in fallback mode (CPU execution without offloading) \\
    \reviewtt{OMP\_TARGET\_CLAUSE\_HAS\_DEVICE\_ADDR(...)} & \reviewtt{has\_device\_addr(\_\_VA\_ARGS\_\_)} & \reviewtt{ACC\_CLAUSE\_DEVICEPTR(\_\_VA\_ARGS\_\_)} & N/A (disregarded in fallback mode) \\
    \reviewtt{OMP\_TARGET\_CLAUSE\_IS\_DEVICE\_PTR(...)} & \reviewtt{is\_device\_ptr(\_\_VA\_ARGS\_\_)} & \reviewtt{ACC\_CLAUSE\_DEVICEPTR(\_\_VA\_ARGS\_\_)} & N/A (disregarded in fallback mode) \\
    \reviewtt{OMP\_TARGET\_CLAUSE\_USES\_ALLOCATORS(...)} & \reviewtt{uses\_allocators(\_\_VA\_ARGS\_\_)} & N/A (disregarded in OpenACC backend) & N/A (disregarded in fallback mode) \\
    \reviewtt{OMP\_TARGET\_CLAUSE\_FROM(...)} & \reviewtt{from(\_\_VA\_ARGS\_\_)} & \reviewtt{ACC\_CLAUSE\_HOST(\_\_VA\_ARGS\_\_)} & N/A (disregarded in fallback mode) \\
    \reviewtt{OMP\_TARGET\_CLAUSE\_TO(...)} & \reviewtt{to(\_\_VA\_ARGS\_\_)} & \reviewtt{ACC\_CLAUSE\_DEVICE(\_\_VA\_ARGS\_\_)} & N/A (disregarded in fallback mode) \\
    \reviewtt{OMP\_TARGET\_PASS\_LIST(...)} & \reviewtt{OMP\_PASS\_LIST(\_\_VA\_ARGS\_\_)} & \reviewtt{ACC\_PASS\_LIST(\_\_VA\_ARGS\_\_)} & \reviewtt{OMP\_PASS\_LIST(\_\_VA\_ARGS\_\_)} \\
    \reviewtt{OMP\_TARGET\_CLAUSE\_SEQ\_CST} & \reviewtt{OMP\_CLAUSE\_SEQ\_CST} & N/A (disregarded in OpenACC backend) & \reviewtt{OMP\_CLAUSE\_SEQ\_CST} \\
    \reviewtt{OMP\_TARGET\_CLAUSE\_ACQ\_REL} & \reviewtt{OMP\_CLAUSE\_ACQ\_REL} & N/A (disregarded in OpenACC backend) & \reviewtt{OMP\_CLAUSE\_ACQ\_REL} \\
    \reviewtt{OMP\_TARGET\_CLAUSE\_RELEASE} & \reviewtt{OMP\_CLAUSE\_RELEASE} & N/A (disregarded in OpenACC backend) & \reviewtt{OMP\_CLAUSE\_RELEASE} \\
    \reviewtt{OMP\_TARGET\_CLAUSE\_ACQUIRE} & \reviewtt{OMP\_CLAUSE\_ACQUIRE} & N/A (disregarded in OpenACC backend) & \reviewtt{OMP\_CLAUSE\_ACQUIRE} \\
    \reviewtt{OMP\_TARGET\_CLAUSE\_RELAXED} & \reviewtt{OMP\_CLAUSE\_RELAXED} & N/A (disregarded in OpenACC backend) & \reviewtt{OMP\_CLAUSE\_RELAXED} \\
    \reviewtt{OMP\_TARGET\_CLAUSE\_READ} & \reviewtt{OMP\_CLAUSE\_READ} & \reviewtt{ACC\_CLAUSE\_READ} & \reviewtt{OMP\_CLAUSE\_READ} \\
    \reviewtt{OMP\_TARGET\_CLAUSE\_WRITE} & \reviewtt{OMP\_CLAUSE\_WRITE} & \reviewtt{ACC\_CLAUSE\_WRITE} & \reviewtt{OMP\_CLAUSE\_WRITE} \\
    \reviewtt{OMP\_TARGET\_CLAUSE\_UPDATE} & \reviewtt{OMP\_CLAUSE\_UPDATE} & \reviewtt{ACC\_CLAUSE\_UPDATE} & \reviewtt{OMP\_CLAUSE\_UPDATE} \\
    \reviewtt{OMP\_TARGET\_CLAUSE\_CAPTURE} & \reviewtt{OMP\_CLAUSE\_CAPTURE} & \reviewtt{ACC\_CLAUSE\_CAPTURE} & \reviewtt{OMP\_CLAUSE\_CAPTURE} \\
    \reviewtt{OMP\_TARGET\_CLAUSE\_COMPARE} & \reviewtt{OMP\_CLAUSE\_COMPARE} & N/A (disregarded in OpenACC backend) & \reviewtt{OMP\_CLAUSE\_COMPARE} \\
    \reviewtt{OMP\_TARGET\_CLAUSE\_FAIL(...)} & \reviewtt{OMP\_CLAUSE\_FAIL(\_\_VA\_ARGS\_\_)} & N/A (disregarded in OpenACC backend) & \reviewtt{OMP\_CLAUSE\_FAIL(\_\_VA\_ARGS\_\_)} \\
    \reviewtt{OMP\_TARGET\_CLAUSE\_WEAK} & \reviewtt{OMP\_CLAUSE\_WEAK} & N/A (disregarded in OpenACC backend) & \reviewtt{OMP\_CLAUSE\_WEAK} \\
    \reviewtt{OMP\_TARGET\_CLAUSE\_HINT(expr)} & \reviewtt{OMP\_CLAUSE\_HINT(expr)} & N/A (disregarded in OpenACC backend) & \reviewtt{OMP\_CLAUSE\_HINT(expr)} \\
    \reviewtt{OMP\_TARGET\_CLAUSE\_SIMD} & \reviewtt{OMP\_CLAUSE\_SIMD} & \reviewtt{ACC\_CLAUSE\_INDEPENDENT} & \reviewtt{OMP\_CLAUSE\_SIMD} \\
    \reviewtt{OMP\_TARGET\_CLAUSE\_DEFAULT\_SHARED} & \reviewtt{OMP\_CLAUSE\_DEFAULT\_SHARED} & N/A (disregarded in OpenACC backend) & \reviewtt{OMP\_CLAUSE\_DEFAULT\_SHARED} \\
    \reviewtt{OMP\_TARGET\_CLAUSE\_DEFAULT\_FIRSTPRIVATE} & \reviewtt{OMP\_CLAUSE\_DEFAULT\_FIRSTPRIVATE} & N/A (disregarded in OpenACC backend) & \reviewtt{OMP\_CLAUSE\_DEFAULT\_FIRSTPRIVATE} \\
    \reviewtt{OMP\_TARGET\_CLAUSE\_DEFAULT\_PRIVATE} & \reviewtt{OMP\_CLAUSE\_DEFAULT\_PRIVATE} & N/A (disregarded in OpenACC backend) & \reviewtt{OMP\_CLAUSE\_DEFAULT\_PRIVATE} \\
    \reviewtt{OMP\_TARGET\_CLAUSE\_DEFAULT\_NONE} & \reviewtt{OMP\_CLAUSE\_DEFAULT\_NONE} & N/A (disregarded in OpenACC backend) & \reviewtt{OMP\_CLAUSE\_DEFAULT\_NONE} \\
    \end{tblr}
\end{sidewaystable*}
\begin{sidewaystable*}[t]
    \footnotesize
    \caption{OpenMP target clauses (3/3)}
    \label{tab:clause:OpenMP.target.3}
    \centering
    \begin{tblr}{
        width={\linewidth},
        colspec={XXXX},
        rowhead=2,
        cells={halign=l},
        hline{1,2,25}={solid},
        vline{2}={solid},
        row{1}={halign=c, font={\bfseries}}
    }
    input & output & counterpart in OpenACC backend & counterpart in fallback mode (CPU execution without offloading) \\
    \reviewtt{OMP\_TARGET\_CLAUSE\_SHARED(...)} & \reviewtt{OMP\_CLAUSE\_SHARED(\_\_VA\_ARGS\_\_)} & N/A (disregarded in OpenACC backend) & \reviewtt{OMP\_CLAUSE\_SHARED(\_\_VA\_ARGS\_\_)} \\
    \reviewtt{OMP\_TARGET\_CLAUSE\_PRIVATE(...)} & \reviewtt{OMP\_CLAUSE\_PRIVATE(\_\_VA\_ARGS\_\_)} & \reviewtt{ACC\_CLAUSE\_PRIVATE(\_\_VA\_ARGS\_\_)} & \reviewtt{OMP\_CLAUSE\_PRIVATE(\_\_VA\_ARGS\_\_)} \\
    \reviewtt{OMP\_TARGET\_CLAUSE\_FIRSTPRIVATE(...)} & \reviewtt{OMP\_CLAUSE\_FIRSTPRIVATE(\_\_VA\_ARGS\_\_)} & \reviewtt{ACC\_CLAUSE\_FIRSTPRIVATE(\_\_VA\_ARGS\_\_)} & \reviewtt{OMP\_CLAUSE\_FIRSTPRIVATE(\_\_VA\_ARGS\_\_)} \\
    \reviewtt{OMP\_TARGET\_CLAUSE\_LASTPRIVATE(...)} & \reviewtt{OMP\_CLAUSE\_LASTPRIVATE(\_\_VA\_ARGS\_\_)} & N/A (disregarded in OpenACC backend) & \reviewtt{OMP\_CLAUSE\_LASTPRIVATE(\_\_VA\_ARGS\_\_)} \\
    \reviewtt{OMP\_TARGET\_CLAUSE\_LINEAR(...)} & \reviewtt{OMP\_CLAUSE\_LINEAR(\_\_VA\_ARGS\_\_)} & N/A (disregarded in OpenACC backend) & \reviewtt{OMP\_CLAUSE\_LINEAR(\_\_VA\_ARGS\_\_)} \\
    \reviewtt{OMP\_TARGET\_CLAUSE\_ALLOCATE(...)} & \reviewtt{OMP\_CLAUSE\_ALLOCATE(\_\_VA\_ARGS\_\_)} & N/A (disregarded in OpenACC backend) & \reviewtt{OMP\_CLAUSE\_ALLOCATE(\_\_VA\_ARGS\_\_)} \\
    \reviewtt{OMP\_TARGET\_CLAUSE\_COLLAPSE(num)} & \reviewtt{OMP\_CLAUSE\_COLLAPSE(num)} & \reviewtt{ACC\_CLAUSE\_COLLAPSE(num)} & \reviewtt{OMP\_CLAUSE\_COLLAPSE(num)} \\
    \reviewtt{OMP\_TARGET\_CLAUSE\_DEPEND(...)} & \reviewtt{OMP\_CLAUSE\_DEPEND(\_\_VA\_ARGS\_\_)} & N/A (disregarded in OpenACC backend) & \reviewtt{OMP\_CLAUSE\_DEPEND(\_\_VA\_ARGS\_\_)} \\
    \reviewtt{OMP\_TARGET\_CLAUSE\_DEPEND\_IN(...)} & \reviewtt{OMP\_CLAUSE\_DEPEND\_IN(\_\_VA\_ARGS\_\_)} & \reviewtt{ACC\_CLAUSE\_WAIT(\_\_VA\_ARGS\_\_)} & \reviewtt{OMP\_CLAUSE\_DEPEND\_IN(\_\_VA\_ARGS\_\_)} \\
    \reviewtt{OMP\_TARGET\_CLAUSE\_DEVICE(...)} & \reviewtt{device(\_\_VA\_ARGS\_\_)} & N/A (disregarded in OpenACC backend) & \reviewtt{device(\_\_VA\_ARGS\_\_)} \\
    \reviewtt{OMP\_TARGET\_CLAUSE\_IF(cond)} & \reviewtt{OMP\_CLAUSE\_IF(cond)} & \reviewtt{ACC\_CLAUSE\_IF(cond)} & \reviewtt{OMP\_CLAUSE\_IF(cond)} \\
    \reviewtt{OMP\_TARGET\_CLAUSE\_IF\_TARGET(cond)} & \reviewtt{OMP\_CLAUSE\_IF(target : cond)} & \reviewtt{ACC\_CLAUSE\_IF(cond)} & \reviewtt{OMP\_CLAUSE\_IF(target : cond)} \\
    \reviewtt{OMP\_TARGET\_CLAUSE\_MAP(...)} & \reviewtt{OMP\_CLAUSE\_MAP(\_\_VA\_ARGS\_\_)} & N/A (disregarded in OpenACC backend) & \reviewtt{OMP\_CLAUSE\_MAP(\_\_VA\_ARGS\_\_)} \\
    \reviewtt{OMP\_TARGET\_CLAUSE\_MAP\_ALLOC(...)} & \reviewtt{OMP\_CLAUSE\_MAP\_ALLOC(\_\_VA\_ARGS\_\_)} & \reviewtt{ACC\_CLAUSE\_CREATE(\_\_VA\_ARGS\_\_)} & \reviewtt{OMP\_CLAUSE\_MAP\_ALLOC(\_\_VA\_ARGS\_\_)} \\
    \reviewtt{OMP\_TARGET\_CLAUSE\_MAP\_TO(...)} & \reviewtt{OMP\_CLAUSE\_MAP\_TO(\_\_VA\_ARGS\_\_)} & \reviewtt{ACC\_CLAUSE\_COPYIN(\_\_VA\_ARGS\_\_)} & \reviewtt{OMP\_CLAUSE\_MAP\_TO(\_\_VA\_ARGS\_\_)} \\
    \reviewtt{OMP\_TARGET\_CLAUSE\_MAP\_FROM(...)} & \reviewtt{OMP\_CLAUSE\_MAP\_FROM(\_\_VA\_ARGS\_\_)} & \reviewtt{ACC\_CLAUSE\_COPYOUT(\_\_VA\_ARGS\_\_)} & \reviewtt{OMP\_CLAUSE\_MAP\_FROM(\_\_VA\_ARGS\_\_)} \\
    \reviewtt{OMP\_TARGET\_CLAUSE\_MAP\_TOFROM(...)} & \reviewtt{OMP\_CLAUSE\_MAP\_TOFROM(\_\_VA\_ARGS\_\_)} & \reviewtt{ACC\_CLAUSE\_COPY(\_\_VA\_ARGS\_\_)} & \reviewtt{OMP\_CLAUSE\_MAP\_TOFROM(\_\_VA\_ARGS\_\_)} \\
    \reviewtt{OMP\_TARGET\_CLAUSE\_MAP\_RELEASE(...)} & \reviewtt{OMP\_CLAUSE\_MAP\_RELEASE(\_\_VA\_ARGS\_\_)} & \reviewtt{ACC\_CLAUSE\_DELETE(\_\_VA\_ARGS\_\_)} & \reviewtt{OMP\_CLAUSE\_MAP\_RELEASE(\_\_VA\_ARGS\_\_)} \\
    \reviewtt{OMP\_TARGET\_CLAUSE\_MAP\_DELETE(...)} & \reviewtt{OMP\_CLAUSE\_MAP\_DELETE(\_\_VA\_ARGS\_\_)} & \reviewtt{ACC\_CLAUSE\_DELETE(\_\_VA\_ARGS\_\_)} & \reviewtt{OMP\_CLAUSE\_MAP\_DELETE(\_\_VA\_ARGS\_\_)} \\
    \reviewtt{OMP\_TARGET\_CLAUSE\_ORDER(...)} & \reviewtt{OMP\_CLAUSE\_ORDER(\_\_VA\_ARGS\_\_)} & N/A (disregarded in OpenACC backend) & \reviewtt{OMP\_CLAUSE\_ORDER(\_\_VA\_ARGS\_\_)} \\
    \reviewtt{OMP\_TARGET\_CLAUSE\_NOWAIT} & \reviewtt{OMP\_CLAUSE\_NOWAIT} & \reviewtt{ACC\_CLAUSE\_ASYNC()} & \reviewtt{OMP\_CLAUSE\_NOWAIT} \\
    \reviewtt{OMP\_TARGET\_CLAUSE\_REDUCTION(...)} & \reviewtt{OMP\_CLAUSE\_REDUCTION(\_\_VA\_ARGS\_\_)} & \reviewtt{ACC\_CLAUSE\_REDUCTION(\_\_VA\_ARGS\_\_)} & \reviewtt{OMP\_CLAUSE\_REDUCTION(\_\_VA\_ARGS\_\_)} \\
    \reviewtt{OMP\_TARGET\_CLAUSE\_IN\_REDUCTION(...)} & \reviewtt{OMP\_CLAUSE\_IN\_REDUCTION(\_\_VA\_ARGS\_\_)} & N/A (disregarded in OpenACC backend) & \reviewtt{OMP\_CLAUSE\_IN\_REDUCTION(\_\_VA\_ARGS\_\_)} \\
    \end{tblr}
\end{sidewaystable*}

\begin{table*}[t]
    \caption{OpenMP clauses (1/2)}
    \label{tab:clause:OpenMP.1}
    \centering
    \begin{tblr}{
        width={\linewidth},
        cells={halign=l},
        hline{1,2,49}={solid},
        vline{2}={solid},
        column{1,2}={font={\ttfamily}},
        row{1}={halign=c, font={\bfseries}}
    }
    input & output \\
    OMP\_CLAUSE\_EXCLUSIVE(...) & exclusive(\_\_VA\_ARGS\_\_) \\
    OMP\_CLAUSE\_INCLUSIVE(...) & inclusive(\_\_VA\_ARGS\_\_) \\
    OMP\_CLAUSE\_ALIGNED(...) & aligned(\_\_VA\_ARGS\_\_) \\
    OMP\_CLAUSE\_INBRANCH & inbranch \\
    OMP\_CLAUSE\_NOTINBRANCH & notinbranch \\
    OMP\_CLAUSE\_SIMDLEN(length) & simdlen(length) \\
    OMP\_CLAUSE\_UNIFORM(...) & uniform(\_\_VA\_ARGS\_\_) \\
    OMP\_CLAUSE\_SIZES(...) & sizes(\_\_VA\_ARGS\_\_) \\
    OMP\_CLAUSE\_FULL & full \\
    OMP\_CLAUSE\_PARTIAL(...) & partial(\_\_VA\_ARGS\_\_) \\
    OMP\_CLAUSE\_COPYIN(...) & copyin(\_\_VA\_ARGS\_\_) \\
    OMP\_CLAUSE\_NUM\_THREADS(num) & num\_threads(num) \\
    OMP\_CLAUSE\_PROC\_BIND(attr) & proc\_bind(attr) \\
    OMP\_CLAUSE\_NUM\_TEAMS(...) & num\_teams(\_\_VA\_ARGS\_\_) \\
    OMP\_CLAUSE\_THREAD\_LIMIT(num) & thread\_limit(num) \\
    OMP\_CLAUSE\_NONTEMPORAL(...) & nontemporal(\_\_VA\_ARGS\_\_) \\
    OMP\_CLAUSE\_SAFELEN(len) & safelen(len) \\
    OMP\_CLAUSE\_FILTER(num) & filter(num) \\
    OMP\_CLAUSE\_COPYPRIVATE(...) & copyprivate(\_\_VA\_ARGS\_\_) \\
    OMP\_CLAUSE\_ORDERED(...) & ordered(\_\_VA\_ARGS\_\_) \\
    OMP\_CLAUSE\_SCHEDULE(...) & schedule(\_\_VA\_ARGS\_\_) \\
    OMP\_CLAUSE\_DIST\_SCHEDULE(...) & dist\_schedule(\_\_VA\_ARGS\_\_) \\
    OMP\_CLAUSE\_BIND(bind) & bind(bind) \\
    OMP\_CLAUSE\_AFFINITY(...) & affinity(\_\_VA\_ARGS\_\_) \\
    OMP\_CLAUSE\_DETACH(eve) & detach(eve) \\
    OMP\_CLAUSE\_FINAL(expr) & final(expr) \\
    OMP\_CLAUSE\_MERGEABLE & mergeable \\
    OMP\_CLAUSE\_PRIORITY(val) & priority(val) \\
    OMP\_CLAUSE\_UNTIED & untied \\
    OMP\_CLAUSE\_GRAINSIZE(...) & grainsize(\_\_VA\_ARGS\_\_) \\
    OMP\_CLAUSE\_NOGROUP & nogroup \\
    OMP\_CLAUSE\_NUM\_TASKS(...) & num\_tasks(\_\_VA\_ARGS\_\_) \\
    OMP\_CLAUSE\_INIT(...) & init(\_\_VA\_ARGS\_\_) \\
    OMP\_CLAUSE\_USE(var) & use(var) \\
    OMP\_CLAUSE\_TASK\_REDUCTION(...) & task\_reduction(\_\_VA\_ARGS\_\_) \\
    OMP\_CLAUSE\_DESTROY(...) & destroy(\_\_VA\_ARGS\_\_) \\
    OMP\_PASS\_LIST(...) & (\_\_VA\_ARGS\_\_) \\
    OMP\_CLAUSE\_SEQ\_CST & seq\_cst \\
    OMP\_CLAUSE\_ACQ\_REL & acq\_rel \\
    OMP\_CLAUSE\_RELEASE & release \\
    OMP\_CLAUSE\_ACQUIRE & acquire \\
    OMP\_CLAUSE\_RELAXED & relaxed \\
    OMP\_CLAUSE\_READ & read \\
    OMP\_CLAUSE\_WRITE & write \\
    OMP\_CLAUSE\_UPDATE & update \\
    OMP\_CLAUSE\_CAPTURE & capture \\
    OMP\_CLAUSE\_COMPARE & compare \\
    \end{tblr}
\end{table*}
\begin{table*}[t]
    \caption{OpenMP clauses (2/2)}
    \label{tab:clause:OpenMP.2}
    \centering
    \begin{tblr}{
        width={\linewidth},
        cells={halign=l},
        hline{1,2,34}={solid},
        vline{2}={solid},
        column{1,2}={font={\ttfamily}},
        row{1}={halign=c, font={\bfseries}}
    }
    input & output \\
    OMP\_CLAUSE\_FAIL(...) & fail(\_\_VA\_ARGS\_\_) \\
    OMP\_CLAUSE\_WEAK & weak \\
    OMP\_CLAUSE\_HINT(expr) & hint(expr) \\
    OMP\_CLAUSE\_THREADS & threads \\
    OMP\_CLAUSE\_SIMD & simd \\
    OMP\_CLAUSE\_DOACROSS(...) & doacross(\_\_VA\_ARGS\_\_) \\
    OMP\_CLAUSE\_DEFAULT(attr) & default(attr) \\
    OMP\_CLAUSE\_DEFAULT\_SHARED & OMP\_CLAUSE\_DEFAULT(shared) \\
    OMP\_CLAUSE\_DEFAULT\_FIRSTPRIVATE & OMP\_CLAUSE\_DEFAULT(firstprivate) \\
    OMP\_CLAUSE\_DEFAULT\_PRIVATE & OMP\_CLAUSE\_DEFAULT(private) \\
    OMP\_CLAUSE\_DEFAULT\_NONE & OMP\_CLAUSE\_DEFAULT(none) \\
    OMP\_CLAUSE\_SHARED(...) & shared(\_\_VA\_ARGS\_\_) \\
    OMP\_CLAUSE\_PRIVATE(...) & private(\_\_VA\_ARGS\_\_) \\
    OMP\_CLAUSE\_FIRSTPRIVATE(...) & firstprivate(\_\_VA\_ARGS\_\_) \\
    OMP\_CLAUSE\_LASTPRIVATE(...) & lastprivate(\_\_VA\_ARGS\_\_) \\
    OMP\_CLAUSE\_LINEAR(...) & linear(\_\_VA\_ARGS\_\_) \\
    OMP\_CLAUSE\_ALLOCATE(...) & allocate(\_\_VA\_ARGS\_\_) \\
    OMP\_CLAUSE\_COLLAPSE(num) & collapse(num) \\
    OMP\_CLAUSE\_DEPEND(...) & depend(\_\_VA\_ARGS\_\_) \\
    OMP\_CLAUSE\_DEPEND\_IN(...) & OMP\_CLAUSE\_DEPEND(in : \_\_VA\_ARGS\_\_) \\
    OMP\_CLAUSE\_IF(...) & if(\_\_VA\_ARGS\_\_) \\
    OMP\_CLAUSE\_MAP(...) & map(\_\_VA\_ARGS\_\_) \\
    OMP\_CLAUSE\_MAP\_ALLOC(...) & OMP\_CLAUSE\_MAP(alloc : \_\_VA\_ARGS\_\_) \\
    OMP\_CLAUSE\_MAP\_TO(...) & OMP\_CLAUSE\_MAP(to : \_\_VA\_ARGS\_\_) \\
    OMP\_CLAUSE\_MAP\_FROM(...) & OMP\_CLAUSE\_MAP(from : \_\_VA\_ARGS\_\_) \\
    OMP\_CLAUSE\_MAP\_TOFROM(...) & OMP\_CLAUSE\_MAP(tofrom : \_\_VA\_ARGS\_\_) \\
    OMP\_CLAUSE\_MAP\_RELEASE(...) & OMP\_CLAUSE\_MAP(release : \_\_VA\_ARGS\_\_) \\
    OMP\_CLAUSE\_MAP\_DELETE(...) & OMP\_CLAUSE\_MAP(delete : \_\_VA\_ARGS\_\_) \\
    OMP\_CLAUSE\_ORDER(...) & order(\_\_VA\_ARGS\_\_ concurrent) \\
    OMP\_CLAUSE\_NOWAIT & nowait \\
    OMP\_CLAUSE\_REDUCTION(...) & reduction(\_\_VA\_ARGS\_\_) \\
    OMP\_CLAUSE\_IN\_REDUCTION(...) & in\_reduction(\_\_VA\_ARGS\_\_) \\
    \end{tblr}
\end{table*}

\section*{Acknowledgment}
This research was partially conducted using Mercury at the Information Technology Center, The University of Tokyo.
This work was supported by JSPS KAKENHI Grant Numbers JP23K11123, JP20K14517, JP20H00580, JP20K21787, and JP22H00507.
This work was also supported by ``Joint Usage/Research Center for Interdisciplinary Large-scale Information Infrastructures (JHPCN)'' and ``High Performance Computing Infrastructure (HPCI)'' in Japan (Project ID: jh240052 and jh240074).

\bibliographystyle{IEEEtran}
\bibliography{ref}

\begin{IEEEbiography}[{\includegraphics[width=1in,height=1.25in,clip,keepaspectratio]{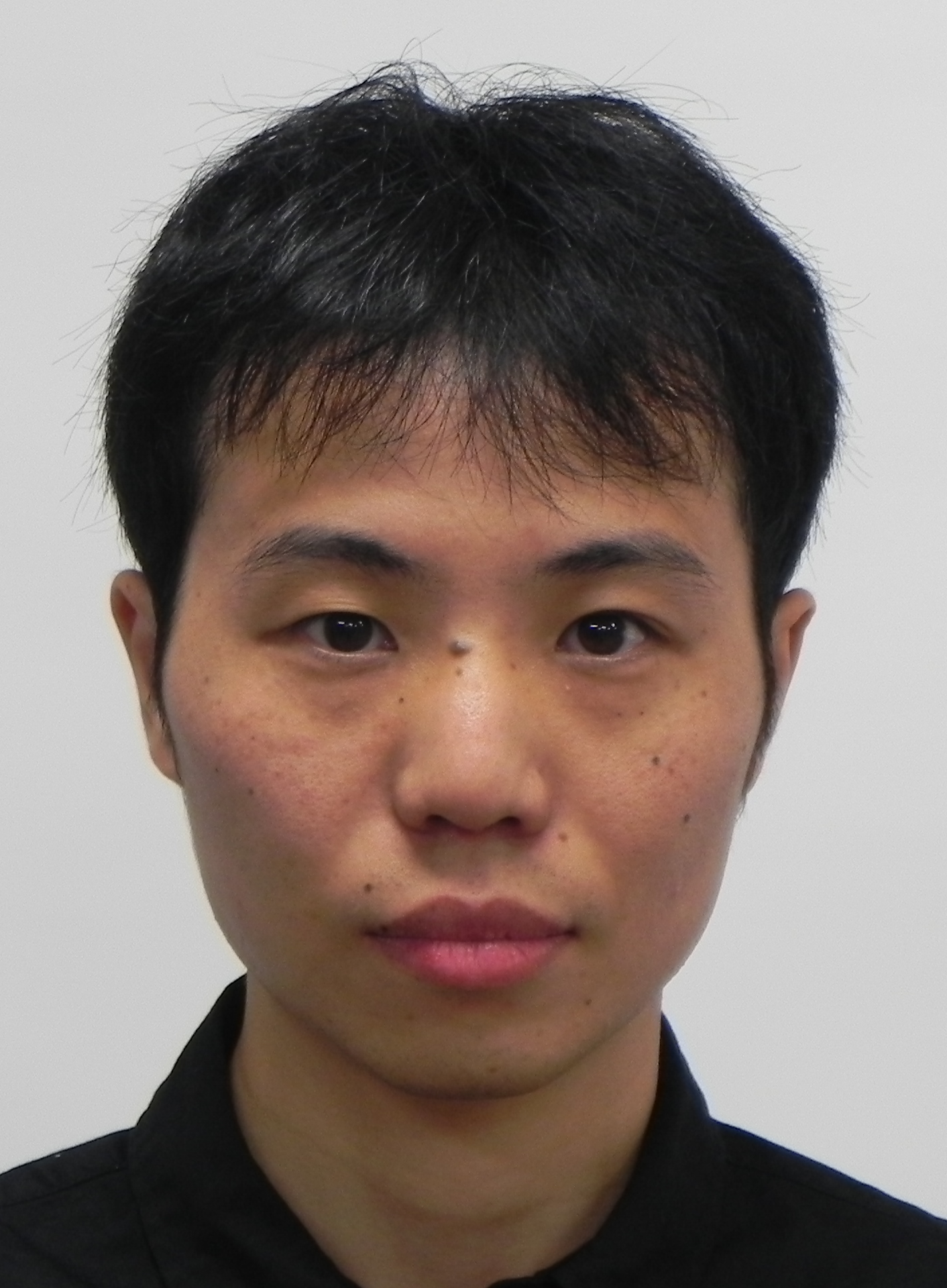}}]{Yohei Miki} (M'23) received the B.S., M.S., and Ph.D. degrees in science from University of Tsukuba, in 2009, 2011, and 2014, respectively, and the M.E. degree in engineering from University of Tsukuba, in 2013.

From 2014 to 2017, he was a Postdoctoral Research Fellow with the Center for Computational Sciences, University of Tsukuba.
From 2017 to 2024, he was a Research Associate at the Information Technology Center, the University of Tokyo, Japan.
Since 2024, he has been an Associate Professor at the Information Technology Center, the University of Tokyo, Japan.
His research interests include formation and evolution of galaxies, coevolution of massive black holes and galaxies, galactic archeology, $N$-body simulations, computational fluid dynamics, high-performance computing, and GPU computing.

Dr. Miki is a member of IAU (International Astronomical Union), ASJ (Astronomical Society of Japan), and IPSJ (Information Processing Society of Japan).
\end{IEEEbiography}

\begin{IEEEbiography}[{\includegraphics[width=1in,height=1.25in,clip,keepaspectratio]{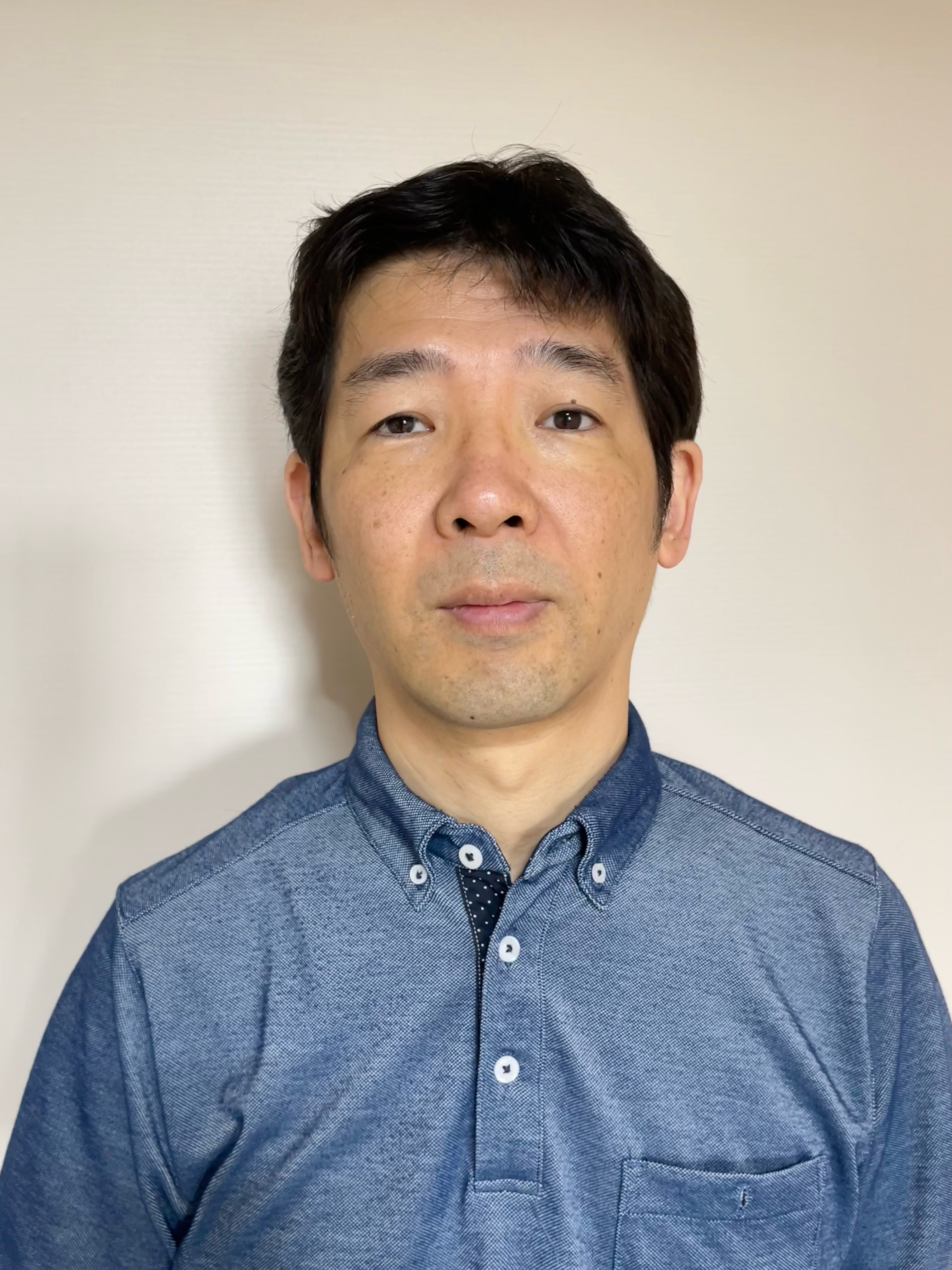}}]{Toshihiro Hanawa} (M'98) received the B.E., M.E., and Ph.D. degrees in engineering from Keio University, Yokohama, in 1993, 1995, and 1998, respectively.

From 1998 to 2007, he was an Assistant Professor with Tokyo University of Technology.
From 2007 to 2013, he was a Research Associate and then an Associate Professor with the Center for Computational Sciences, University of Tsukuba.
From 2013 to 2020, he was an Associate Professor with the Information Technology Center, the University of Tokyo.
Since 2020, he has been a Professor with the Information Technology Center, the University of Tokyo.
His research interests include high-performance computing, GPU computing, interconnect, and FPGA computing.
\end{IEEEbiography}

\EOD

\end{document}